\documentclass[a4paper, amsfonts, amssymb, amsmath, reprint, showkeys, nofootinbib, twoside, superscriptaddress]{revtex4-1}
\usepackage[english]{babel}
\usepackage[utf8]{inputenc}
\usepackage[colorinlistoftodos, color=green!40, prependcaption]{todonotes}
\usepackage{amsthm}
\usepackage{amsmath}
\usepackage{mathtools}
\usepackage{physics}
\usepackage{xcolor}
\usepackage{graphicx}
\usepackage[left=23mm,right=13mm,top=35mm,columnsep=15pt]{geometry} 
\usepackage{adjustbox}
\usepackage{placeins}
\usepackage[T1]{fontenc}
\usepackage{lipsum}
\usepackage{csquotes}
\usepackage{siunitx}
\usepackage{upgreek}

\usepackage[caption=false]{subfig}

\usepackage{comment}
\usepackage{enumitem}
\usepackage{amsmath,bm}

\usepackage[pdftitle={Article}, pdfauthor={Author}, colorlinks=true]{hyperref}

\begin{document}
\title{Dephasing due to electromagnetic interactions in spatial qubits }

\author{Martine Schut}
    \affiliation{Van Swinderen Institute for Particle Physics and Gravity, University of Groningen, 9747 AG Groningen, the Netherlands }
    \affiliation{Bernoulli Institute for Mathematics, Computer Science and Artificial Intelligence, University of Groningen, 9747 AG Groningen, the Netherlands \vspace{1mm}}

\author{Herre Bosma}
    \affiliation{Van Swinderen Institute for Particle Physics and Gravity, University of Groningen, 9747 AG Groningen, the Netherlands }

    \author{MengZhi Wu}
    \affiliation{Van Swinderen Institute for Particle Physics and Gravity, University of Groningen, 9747 AG Groningen, the Netherlands }

\author{Marko Toro\v{s}}
    \affiliation{School of Physics and Astronomy, University of Glasgow, Glasgow, G12 8QQ, UK}
\author{Sougato Bose}
    \affiliation{Department of Physics and Astronomy, University College London, Gower Street, WC1E 6BT London, UK}
\author{Anupam  Mazumdar}
    \affiliation{Van Swinderen Institute for Particle Physics and Gravity, University of Groningen, 9747 AG Groningen, the Netherlands }


\begin{abstract}
Matter-wave interferometers with micro-particles will enable the next generation of quantum sensors to probe minute quantum phase information.
Therefore, estimating the loss of coherence and the degree of entanglement degradation for such interferometers is essential. 
In this paper, we will provide a noise analysis in frequency-space focusing on electromagnetic sources of dephasing.
We will assume that our matter-wave interferometer has a residual charge or dipole which can interact with a neighbouring particle in the ambience. We will investigate the dephasing due to the Coulomb, charge-induced dipole, charge-permanent dipole, and dipole-dipole interactions.
All these interactions constitute electromagnetically driven dephasing channels that can affect single or multiple interferometers. As an example, we will apply the obtained formulae to situations with two adjacent micro-particles, which can provide insight for the noise analysis in the quantum gravity-induced entanglement of masses (QGEM) protocol and the C-NOT gate: we will compute the dephasing due to a gas of environmental particles interacting via dipole-dipole and charge-charge couplings, respectively. To obtain simple analytical dephasing formulae, we will employ uniform probability distributions for the impact parameter and for the angles characterizing the relative orientation with respect to the interferometer and a Gaussian distribution for the velocities of the environmental particles. In both cases, we will show that the dephasing rate grows with the number density of particles present in the vacuum chamber, as expected.


\end{abstract}

\maketitle
\tableofcontents
\section{Introduction}\label{sec:introduction}

One of the key features of quantum mechanics is the matter-wave duality exhibited in interferometry with massive systems~\cite{de1923waves}. Matter-wave interferometry has played a central role in many experimental breakthroughs in quantum mechanics~\cite{thomson1927diffraction,davisson1927diffraction,davisson1928reflection}, and also illustrates the idea of the spatial quantum superposition~\cite{Schrodinger:1935zz}.  Matter-wave interferometry has been used to detect the Earth's gravitationally-induced phase in a series of seminal experiments with neutrons and atoms~\cite{colella1975observation,nesvizhevsky2002quantum,fixler2007atom,asenbaum2017phase,overstreet2022observation}. Moreover, it has been suggested as a tool for sensing gravitational waves~\cite{Marshman:2018upe}, neutrinos~\cite{Kilian:2022kgm}, and as a probe for physics beyond the standard model~\cite{Barker:2022mdz}.

When two or more interferometers can be placed adjacent to each other they can also test the quantum entanglement feature~\cite{einstein1935can,bell1964einstein,Horodecki:2009zz}. Recently, it has been suggested that matter-wave interferometry with micro-particles will be sensitive enough to probe the quantum gravity-induced entanglement of masses (QGEM)~\cite{Bose:2017nin,ICTS}, see also~\cite{Marletto:2017kzi}. The entanglement between two adjacent matter-wave interferometers will be observed if gravity is a quantum mechanical entity. In contrast, no entanglement will be generated if gravity is inherently classical as formalized by the local operations and classical communication (LOCC) theorem ~\cite{Bennett:1996gf,Bose:2017nin,Marshman:2019sne,Bose:2022uxe}. Such entangled pairs also provide the basis for a quantum computer as they form the controlled-NOT (CNOT) gate, or the Molmer-Sorensen gate in the context of an ion trap~\cite{Soerensen:1998mp}. One can go one step further and test the quantum light bending interaction~\cite{Biswas:2022qto}, the quantum version of the equivalence principle~\cite{Bose:2022czr,Chakraborty:2023kel}, test for massive graviton~\cite{Elahi:2023ozf}, non-local aspects of quantum gravity~\cite{Vinckers:2023grv}, and
verify the quantum nature of gravity in the process of
measurement~\cite{Hanif:2023fto}. 

Given so many applications on the horizon of the next generation of matter-wave interferometers, it is essential to understand various causes of decoherence and dephasing. Large spatial quantum superpositions~\cite{vandeKamp:2020rqh,PhysRevLett.111.180403,PhysRevLett.117.143003,Margalit:2020qcy,Marshman:2021wyk,Folman2013,folman2019,Folman2018,Zhou:2022frl,Zhou:2022jug,Zhou:2022epb,Marshman:2023nkh} have to remain coherent for the duration of the experiment to extract their delicate experimental signature~\cite{Schut:2021svd,Tilly:2021qef,PhysRevLett.125.023602,Rijavec:2020qxd,RomeroIsart2011LargeQS,Chang_2009,Schlosshauer:2014pgr,bassi2013models,Toros:2020krn,Fragolino:2023agd}. Here we will focus on electromagnetic sources of dephasing induced by ambient particles located in the vicinity of the interferometeric setup.

First, we will consider a single matter-wave interferometer. We will investigate the acceleration noise caused by a single particle passing by the interferometer and estimate how it will dephase the matter-wave interferometer via electromagnetic interactions. 
Then, we will consider if two such entangled interferometers were kept adjacent; the source of acceleration noise will also contribute to entanglement degradation.
If the matter-wave interferometer consists of a neutral micro-particle, the dominant contribution to the acceleration results from its dipole (either permanent or induced). 
If the microparticle possesses any residual charge, such a charged interferometer can interact with an ambient charged particle or an ambient neutral particle possessing a dipole (either permanent or induced). 
The common aspect of all these interactions is that the moving particle will always create a slight jitter in the paths of the interferometer introducing noise and thus dephasing. 
We will analyse these cases systematically and quantify each scenario's resulting loss of coherence.

To analyze the electromagnetic acceleration noise and dephasing we will employ the frequency space techniques, which are commonly used to investigate Newtonian noise~\cite{Saulson:1984}. The reason is that the Coulomb and Newtonian interactions are long-range and provide a $1/r$ potential, resulting in an infinite total cross-section unless a cutoff for small scattering angles is applied~\cite{griffiths_introduction_2018}. Since the computations of the decoherence rate based on the scattering theory generally depend on the total cross-section, see~\cite{Schlosshauer:2014pgr,Schlosshauer:2019ewh}, at least for long-range interactions, we have to use an alternative methodology. In this paper, we will adapt the methodology based on the Feynman path-integral approach following Storey and Cohen-Tannoudji~\cite{Storey:1994oka}, which we have previously used to investigate non-inertial and gravity gradient noise~\cite{Toros:2020dbf,Wu:2022rdv}, to the case of electromagnetic interactions. 

We will begin the paper with a generic discussion on relative acceleration noise (Sec.~\ref{sec:RAN}), which we will apply to electromagnetic noise sources. We will discuss the case of a charged and a neutral interferometer (Sec.~\ref{sec:setup}). First, the interaction between the charged interferometer and an external charge (Coulomb interaction), and the interaction between the charged interferometer and an induced or permanent external dipole (charge-dipole interaction) (Sec.~\ref{sec:charged}). Second, we will analyze the neutral interferometer particle with a permanent or induced dipole that interacts with an external charge (dipole-charge interaction) or with an external dipole (dipole-dipole interaction) (Sec.~\ref{sec:neutral}).

\section{Relative Acceleration Noise}\label{sec:RAN}

This section introduces the main tools for describing relative acceleration noise (RAN) in frequency space. We will then discuss how to compute the resulting dephasing from the associated phase noise.
In a nutshell, any movement of charges or dipoles in the vicinity of the matter-wave interferometer will cause a tiny jitter/acceleration noise, which we denote here by $a_\text{noise, EM}$, which will induce phase fluctuations and hence dephasing. The subscript $\text{EM}$ denotes electromagnetic-type external interactions.  

We will assume that the spatial superposition state can be created via the Stern-Gerlach (SG) setup, as it happens in spin-embedded systems with a nitrogen vacancy (NV). 
We can envisage a spin-magnetic field interaction responsible for creating the superposition, e.g. using an SG apparatus, similar to what has been applied to the atomic case~\cite{Folman2013,folman2019, Margalit:2020qcy,Keil2021,Amit_2019}, 
and charged case~\cite{Henkel_2019}.  The other possibility will be to create a spatial superposition in an ion trap~\cite{Wineland:1992,Wineland:1995,kielpinski2001recent}. In either case, the matter-wave interferometer will be considered a spatial qubit, where interactions with an environment can induce a relative phase between the interferometer's two arms (or the spatial superposition states of the spatial qubit).



\subsection{General noise analysis}\label{subsec:generic_noise}

The difference in the phase picked up between the two arms of an interferometer is determined by the difference in the actions of the two trajectories. 
Taking the superposition to be along one dimension, e.g., in the $\hat{x}$-direction, without any loss of generality, it follows that the difference in phase-shift is given by~\cite{Storey:1994oka}:
\begin{equation}\label{eq:delta-phi-def}
    \begin{aligned}
    \delta \phi & = \phi_R - \phi_L \\
    & = \frac{1}{\hbar}\int_{t_i}^{t_f} L_{R}\left[x_{R},\dot{x_{R}}\right] - L_{L}\left[x_{L}, \dot{x_L}\right] dt,
    \end{aligned}
\end{equation}
where $\phi_i$ and $L_i$ are the phases and the Lagrangians for $i=L, R$, the left and the right arm of the superposition, respectively. Here, we consider a single interferometer; we will consider the case of adjacent interferometers later. The bounds of the integral are equal to the time of splitting and recombination of the two trajectories to complete a one-loop interferometer. The time when the interferometer is in a spatial superposition is given by $\tau= t_f- t_i$, where $f, i$ stands for the interferometer's final and initial time. 

Here, we consider the non-relativistic regime of a free-falling test mass split into the superposition by an acceleration, $\lambda_j$, in the $\hat{x}$-direction~\footnote{The acceleration of the two superposition states is due to the SG setup in the context of NV centred micro-particles~\cite{Bose:2017nin,Marshman:2021wyk,WoodPRA22_GM}. 
In the case of ions, the force exerted to create the superposition will be due to the photon kicks~\cite{Wineland:1992,Wineland:1995}. We can also envisage creating quantum superposition for the charged micro-spheres with the help of SG setup, see~\cite{Henkel_2019}.}.
The Lagrangian of the two arms of the interferometer (with mass $m$) can be expressed as:
\begin{equation}\label{eq:lagrangian}
    L_j = \frac{1}{2}mv_j^2 - m \lambda_j(t) x_j - m \,a_{\operatorname{noise,EM}}(t) x_j\,.
\end{equation}
We wish to close the trajectories (recombine the spatial qubit into a spin qubit) so that the wavefunctions of both trajectories nearly overlap. If they do not, there will be a substantial loss of visibility. 
If the spread of the wavepackets is $\sigma$ then one would require $|X_{L}(t_f )- X_{R}(t_f )|\ll \sigma$ to achieve the required visibility~\cite{Schwinger,Scully,Englert}.

The noise we will consider in the trajectories can be modelled by the fluctuations in the phase, given by the last term in Eq.~\eqref{eq:lagrangian}. 
We assume that $a_{\rm{noise, EM}}$ is independent of the force responsible for creating the superposition but that it is time-dependent during $\tau =t_{f}-t_i$. 
Hence, the fluctuation in the phase shift at the leading order due to the EM interaction will be given by:
\begin{equation}\label{eq:phase-shift}
    \delta \phi = \frac{m}{\hbar} \int_{t_i}^{t_f} a_{\operatorname{noise,EM}}(t) \left(x_R-x_L\right)dt ,
\end{equation}
here $\delta \phi(t)$ will be treated as a statistical quantity.
The measurable noise in any experiment is the statistical average of any stochastic entity, which we will denote by $\mathbb{E}[\cdot]$. 
The averaging can be obtained over time using a single realization of the noise for a time-varying ergodic noise. For example, the average of a time-varying stochastic quantity $\delta \phi(t)$ can be expressed as:
\begin{equation}\label{eq:Teq}
\mathbb{E}[\delta \phi]=\frac{1}{T}\int_0^{T}\delta \phi(t) dt\,,
\end{equation}
where $T$ is much larger than any time scale characterizing the noise. 
For simplicity of the analysis, we will assume that the interferometric time scale is much shorter than the time scale $T$, i.e. $T \gg \tau$, where $\tau$ denotes the experimental time, $\tau = t_f -t_i$.
Furthermore, we will take $T$ as the total time of the experiment (comprising repeated runs of the experimental sequence, where the statistic is gathered).
For concreteness, we will set $T=b/v$ such that in the total experimental time, we sufficiently capture the experimental signature of the external EM source on the experimental statistics following Ref.~\cite{Saulson:1984}.

While the average of the noise can often be assumed to be zero, the autocorrelation of the acceleration noise measured at different times, 
$\mathbb{E}[a_\text{noise, EM}(t_1), a_\text{noise, EM}(t_2)]$, is often non-zero. It is related to the power spectrum density (PSD) of the noise, denoted by $S_{aa}(\omega)$.
The definition of the PSD of the noise $a_\text{noise, EM}$ is 
\begin{equation}\label{eq:saa_def}
S_{aa}(\omega) = \lim\limits_{T\to\infty}\frac{|a_\text{noise, EM}(\omega)|^2}{T},
\end{equation}
where $a_\text{noise, EM}(\omega)$ is the Fourier transform of $a_\text{noise, EM}(t)$ on the time domain $[0,T]$.
According to the Wiener-Khinchin theorem,  the autocorrelation function of a  random process is given by the PSD of that process~\cite{Wiener:1930,Khintchine1934}:
\begin{equation}\label{eq:PSD_wienerkhinchin}
\begin{aligned}
    \mathbb{E}\big[a_{\operatorname{noise,EM}}(t_1)&a_{\operatorname{noise,EM}}(t_2)\big] \\
    &= \frac{1}{2\pi}\int_{\omega_{\rm min}}^{\infty} S_{aa}(\omega)e^{i\omega(t_1-t_2)} d\omega \, ,
\end{aligned}
\end{equation}
where the minimum frequency $\omega_{\operatorname{min}} = 2\pi/(t_f-t_i)$ is the frequency resolution in the experiment~\footnote{
For a real-valued random process $X(t)$, the PSD is defined as:
\begin{align}
    S_{X}(\omega)
    = \lim_{T\to\infty} \frac{1}{T} \int_{-\frac{T}{2}}^{\frac{T}{2}}\int_{-\frac{T}{2}}^{\frac{T}{2}} \mathbb{E}\big[X(t_1)X(t_2)\big] 
    e^{-i \omega (t_1 - t_2)} \dd{t_2}\dd{t_1}. \label{eq:sx}
\end{align}
Assuming that the PSD is stationary (i.e., its properties such as mean and covariance do not change over time) and using the Wiener-Khinchin theorem, we find from Eq.~\eqref{eq:sx}:
\begin{align}
    S_{X}(\omega)
    = \int_{-\infty}^{\infty} \mathbb{E}\big[X(t)X(t+\tau)\big]  e^{-i \omega \tau} \dd{\tau}.
\end{align}
where $\tau=t_2-t_1$, and the result does not depend on the chosen value of $t$ (e.g., we can set $t=0$). Taking the Fourier transform gives eq.~\eqref{eq:PSD_wienerkhinchin}.
}.
In particular, signals and noises are sampled as a discrete finite data series, so the frequency-domain analysis has to be based on the discrete Fourier transform, which has such a frequency resolution~\cite{Chatfield:2016}.

Using the relation in Eq.~\eqref{eq:PSD_wienerkhinchin} and the difference in phase-shift form Eq.~\eqref{eq:phase-shift}, the variance $\mathbb{E}[(\delta \phi)^2]$ can be written in the Fourier space by:
\begin{equation}\label{eq:GammaDef}
    \Gamma_{\operatorname{n}} = \mathbb{E}\left[(\delta \phi)^2\right] = \frac{1}{2\pi} \left(\frac{m}{\hbar}\right)^2 \int_{\omega_{\rm min}}^{\infty} d\omega S_{\operatorname{aa}}(\omega)F(\omega)\,.
\end{equation}
where $\Gamma_{\operatorname{n}}$ in Eq.~\eqref{eq:GammaDef} is the variance in the noise~\footnote{
Although the term \textit{noise} often signifies \textit{unknown dynamics}, here we use it more loosely  when referring to \textit{unwanted interactions} with environmental particles which we cannot measure directly. The discussion is thus far generic and could span a number of different models of the environment. In later sections we will assume specific probability distributions characterizing the environmental particles (i.e., a dilute gas of particles), and then compute the resulting dephasing after averaging over the distributions. }, which is an observable entity, and $F(\omega)$ is known as  the transfer function~\cite{Wu:2022rdv,Toros:2020dbf}.  As shown below, $\Gamma_{\rm n}$ will be useful to estimate the dephasing for the two spatial qubits, adding up to any other sources of decoherence caused by the external/internal degrees of freedom. Note that the $\Gamma_n$ is a dimensionless entity.

The transfer function is equal to the absolute value squared of the Fourier transform of the difference in the trajectories:
\begin{equation}
\begin{aligned}
    F(\omega) &= \left|\int_{t_i}^{t_f}dt\left(x_L-x_R\right)e^{i\omega t}\right|^2 \\
    &= \left|- \frac{1}{\omega^2}\int_{t_i}^{t_{f}}\frac{d^2}{dt^2}\left(x_L -x_R\right)e^{i\omega t}dt \right|^2,
\end{aligned}
\end{equation}
In the second line, we find the dependency on the acceleration, $\text{d}^2 x/\text{d}t^2$, which, as we will see, can be related to particle equations of motion, and follows from integration by parts and the equality of both the position and velocity at the path's endpoints from the first line.

The dephasing formula in Eq.~\eqref{eq:GammaDef} is still generic and applies to an arbitrary acceleration noise PSD $S_{aa}(\omega)$ and transfer function $F(\omega)$. In the next sections, we will apply it to investigate specific cases of electromagnetic interactions, where the environment is made of gas particles characterized by probability distributions for the impact factor, velocities, and the relative angles with respect to the interferometer. The acceleration noise $S_{aa}(\omega)$ could be either obtained from independent measurements characterizing the environment of the experiment or by developing a microscopic modelling of the environment. To obtain simple formulae we will consider a simple microscopic model of a dilute gas, where the particles move on straight lines, and the interaction among the environmental particles is neglected. The transfer function $F(\omega)$ is on the other hand determined by the interferometer and the type of interaction. In the next section we specify the form of $F(\omega)$ by assuming an idealized Stern-Gerlach interferometer.

\subsection{Example of Stern-Gerlach interferometer}

The dephasing in Eq.~\eqref{eq:GammaDef} is dependent on the specific trajectory of the two arms of the matter-wave interferometer (via the transfer function $F(\omega)$). Here, we assume an SG-type interferometer that is often proposed in the context of entanglement-based quantum-gravity experiments, see~\cite{Bose:2017nin,Marshman:2019sne,Zhou:2022jug,Zhou:2022frl,Zhou:2022epb,Japha:2022xyg}. 

\begin{figure}
    \centering
    \includegraphics[width=0.49\textwidth]{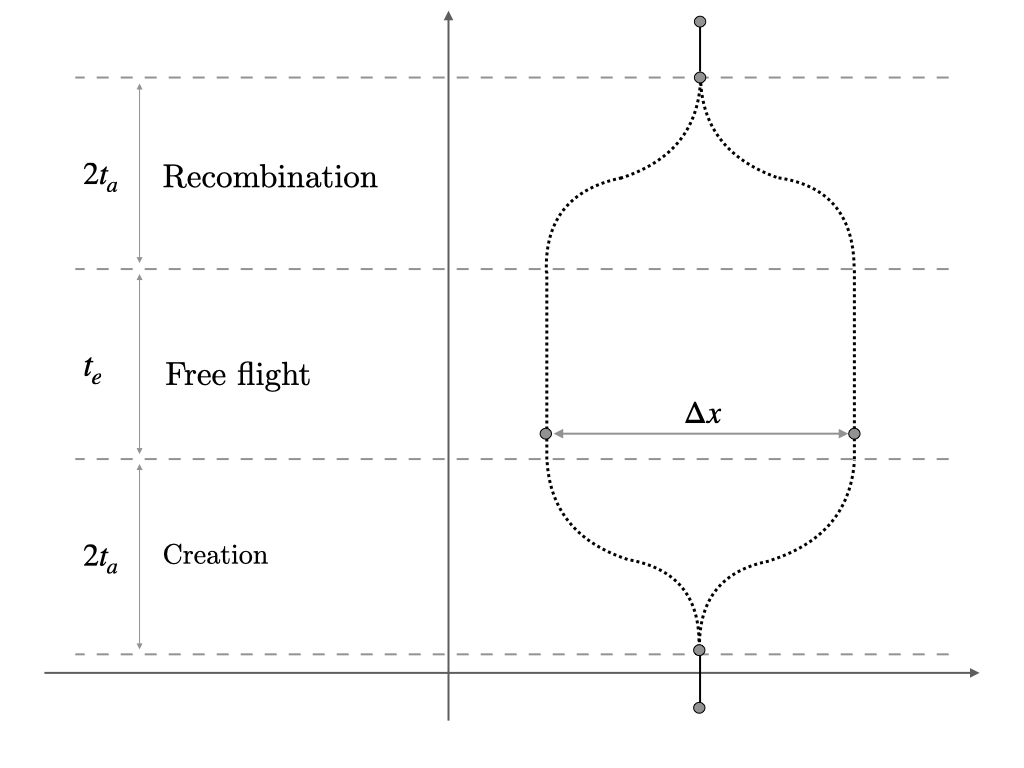}
    \caption{Schematic illustration of the paths of the arms of the interferometer (adapted from~\cite{Toros:2020dbf,Wu:2022rdv}). 
    Based on this figure, the transfer function in Eq.~\eqref{eq:transferfunction2} can be calculated. 
    For the left arm, the acceleration is in the negative direction from $[0, t_a]$ and $[3t_a+t_e, 4t_a+t_e]$, in the positive direction from $[t_a, 2t_a]$ and $[2t_a+t_e, 3t_a+t_e]$, while it is zero for the time range $[2t_a, 2t_a+t_e]$.
    Meanwhile, the acceleration is the opposite for the right arm.}
    \label{fig:Interferometer path}
\end{figure}

We specify the trajectory of two arms of the interferometer; see Fig.~\ref{fig:Interferometer path}.
Here, we assume a symmetric path for the interferometer as a prototype model.
We consider a simplistic scenario, assuming that the acceleration of the spin state $j$, denoted $\lambda_j$, is proportional to the gradient of the magnetic field, given by:
\begin{equation}
    \lambda_j = s_j \frac{g\mu_B}{m}|\nabla B|
\end{equation}
where $s_j$ is the spin of the state, $g$ is the Lande g-factor of the electron ($g=2$), $\mu_B = 9\cdot 10^{-24}\operatorname{J\,T}^{-1}$ is the Bohr-magneton, $m$ is the mass of the test-mass and $|\nabla B|$ is the magnitude of the gradient of the magnetic field. 
To create the path in Fig.~\ref{fig:Interferometer path}, one controls the spin so that the acceleration of the right arms is positive from $[0, t_a]$ and $[3t_a+t_e, 4t_a+t_e]$ and negative from $[t_a, 2t_a]$ and $[2t_a+t_e, 3t_a+t_e]$, while it is zero for the time range $[2t_a, 2t_a+t_e]$. 
For the left arm, the spin and acceleration will be equal in magnitude but in the opposite direction. 
We also assume  $x_j(0) = 0$ and $\dot x_j(0) = 0$ for solving the equations of the motion for the trajectory, given by: $\ddot x_{j}(t)=\lambda_j(t)$.
Micro-particles can be electrically charged by ultraviolet irradiation~\cite{Frimmer_2017}. For a small number of charges $n$, the Lorentz force acting on the micro-particle $\sim nevB/m$ (with $e$ the unit of electric charge) is negligible due to its large mass $m$, see~\cite{Henkel_2019}. 
With the help of the equations of motion, the transfer function corresponding to the trajectory illustrated in Fig.~\ref{fig:Interferometer path} can be computed:
\begin{equation}\label{eq:transferfunction2}
    \begin{aligned}
    F(\omega) &= \frac{1}{\omega^2}\left|\int_{t_i}^{t_{f}}\left(\lambda_L -\lambda_R \right)e^{i\omega t}dt \right|^2
     = \left|\frac{2}{\omega^2}\int_{t_i}^{t_{f}} \lambda_s e^{i\omega t}dt \right|^2\\
    & = \frac{32 \lambda_s^2}{\omega^6} \sin ^4\left(\frac{t_a \omega}{2}\right) \sin ^2\left(\frac{\omega}{2}(2t_a+t_e)\right) \, .
    \end{aligned}
\end{equation}
Since how the superposition is created is immaterial, for the calculation of the transfer function, it is convenient to define the transfer function in terms of the maximal superposition size created after two acceleration times, $\Delta x=\lambda t_a^2$:
\begin{equation}\label{eq:transferfunction}
    F(\omega) = \frac{32 (\Delta x)^2}{\omega^6 t_a^4} \sin ^4\left(\frac{t_a \omega}{2}\right) \sin ^2\left(\frac{\omega}{2}(2t_a+t_e)\right).
\end{equation}
By finding the PSD, $S_{aa}$, due to different types of interaction and combining it with the specifics of the trajectory encoded in the transfer function in Eq.~\eqref{eq:transferfunction}, we can find the noise via Eq.~\eqref{eq:GammaDef}.

\section{EM induced acceleration noise}\label{sec:setup}

In this section, we investigate the situation of external particles interacting electromagnetically with our interferometer, causing spurious accelerations.  We envisage a situation of a rarefied gas, where the external particles can be considered as independent sources of disturbance moving at constant velocity. More refined models of the gaseous environment (e.g., which would include interactions among environmental particles, etc.) could be obtained by refining the model of $S_{aa}(\omega)$, which will be presented below.

\begin{figure}[t]
    \centering
    \includegraphics[width=0.49\textwidth]{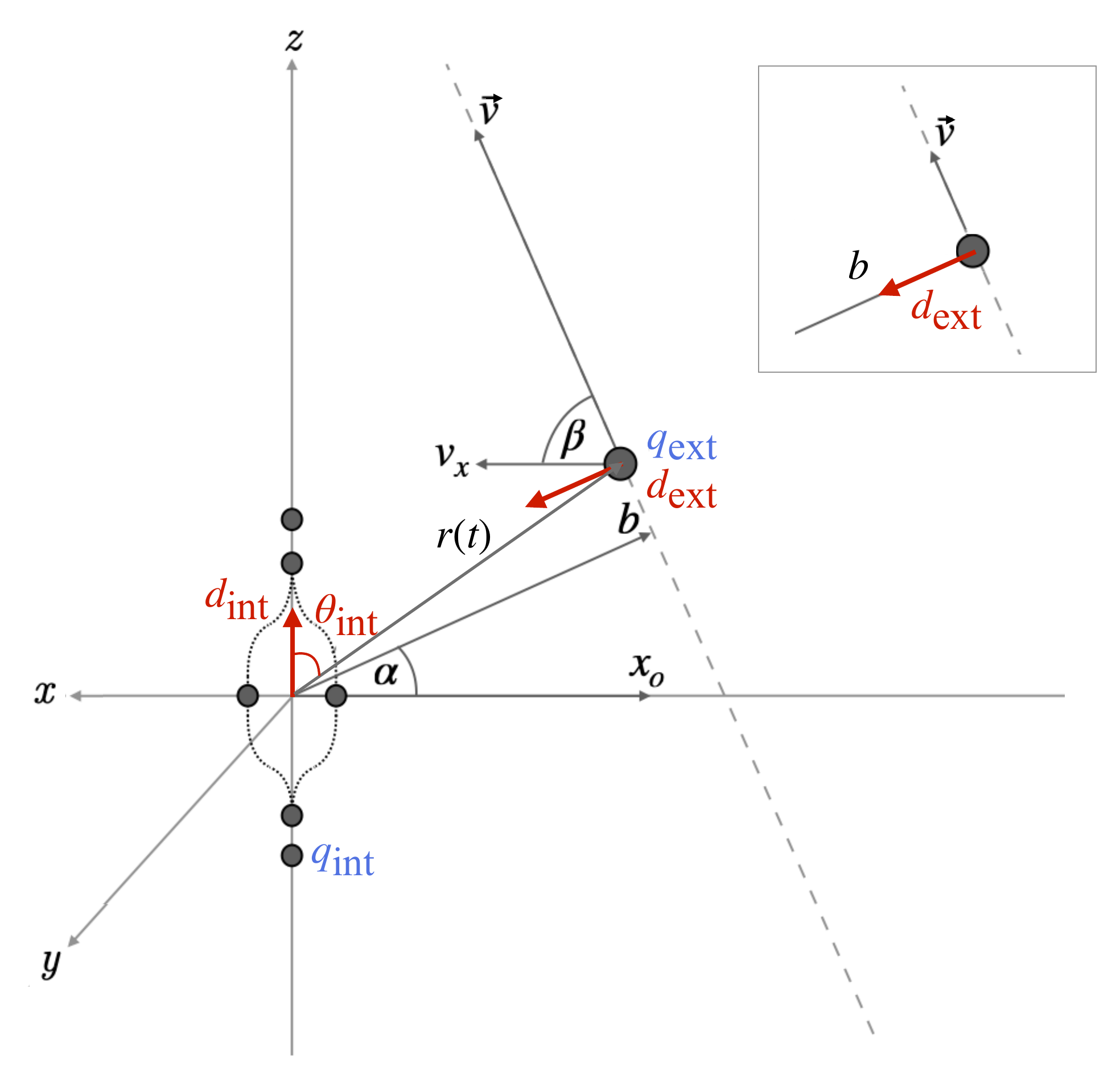}
    \caption{Illustration of the interferometer and the external particle (i.e., the source of dephasing).  The position of the closest approach is $(x_0,y_0,z_0)$ corresponding to the impact parameter $b$ (the centre-of-mass of the interferometer is used for the definition since we assume $\Delta x \ll b$).  The magnitude of the position vector is given by $|\bm{r}|=\sqrt{b^2+(vt^2)}$, where the velocity is $v=|\bm{v}|=(v_x,v_y,v_z)$. The projection angles are defined by $\cos(\alpha)=x_0/b$\; and\; $\cos(\beta)=v_x/v$. The red arrows denote the dipole moments. }
    \label{fig:external particle}
\end{figure}

We will assume that the individual external particle is moving with a constant velocity $\bm{v}$ and with an impact parameter $b$ for the matter-wave interferometer (see Fig.~\ref{fig:external particle}). The position of the external particle concerning the matter-wave interferometer can be conveniently described by the vector $\boldsymbol{r}(t)$ from the interferometer centre of mass to the external particle, with length:
\begin{equation}\label{eq:distancevector}
    r(t)=\sqrt{b^2+(vt)^2}.
\end{equation}
The $x$-component of this vector, from the properties of the setup (see Fig.~\ref{fig:external particle}), is $r_x(t) = x_0 + v_x t$.
The unit vector in the $x$-direction is denoted by  $\hat{\boldsymbol{r}}_x$, and $x_0$ and $v_x$ are defined to be the $x$-components of the impact parameter $b$ and velocity $\bm{v}$, respectively. We can thus introduce the projection angles: 
\begin{equation}\label{eq:projectionangles}
    \cos(\alpha)=x_0/b, \qq{}
    \cos(\beta)=v_x/v.
\end{equation}

The acceleration PSD in the considered situation is thus parameterized by four parameters
\begin{equation}
    S_{aa}(\omega)\equiv S_{aa}(b,v,\alpha,\beta;\omega).
\end{equation}
As these parameters are unknown, we can treat them as random variables characterized by their respective probability distributions. The averaged acceleration noise PSD is given by
\begin{alignat}{2}\label{eq:Saa2}
    \bar{S}_{aa}(\omega) = N \int & db \int dv \int d\alpha \int d\beta \,\, p_\text{b}(b) \,\, p_\text{v}(v)  \nonumber\\
    & p_\upalpha(\alpha) \,\, p_{\upbeta}(\beta) \,\, S_{aa}(b,v,\alpha,\beta;\omega),
\end{alignat}
where $N$ denotes the number of external particles in the experimental chamber, $p_\text{b}(b)$ is the distribution of the impact factors, $p_\text{v}(v)$ is the velocity distribution, and $p_\upalpha(\alpha)$ and $p_{\upbeta}(\beta)$ are the distributions of the relative orientations. We suppose that the probability distributions are given by:~\footnote{
The distribution of the velocity is taken to be Maxwell-Boltzmann.
The distributions for the projection angles and the impact parameter are uniform.
The integrations over the projection angles are divided by the total angle $2\pi$ to get an average angle.
The distribution for the impact parameter is taken to be uniform in Cartesian coordinates and rewritten in terms of spherical coordinates.
}
\begin{alignat}{2}
p_\text{b}(b)&=\frac{3 b^2}{L^3},\,\, b \in   [R,L],\label{eq:1}  \\ 
p_\text{v}(v)&=  \left[\frac{m_\text{gas}}{2\pi k_B T_\text{gas}}\right]^\frac{3}{2} 4\pi v^2 e^{-\frac{m_\text{gas} v^2}{2k_B T_\text{gas}}},\,\, v \in [0,\infty],  \label{eq:MBD}\\ 
p_\upalpha(\alpha)&=\frac{1}{2\pi},\,\,\alpha \in [0,2\pi], \\ 
p_{\upbeta}(\beta)&=\frac{1}{2\pi},\,\,\beta \in [0,2\pi] \label{eq:4},
\end{alignat}
where  $L$ denotes the size of the experimental chamber, $R$ is the radius of the test mass, $m_\text{gas}$ is the mass of the gas particle, and $T_\text{gas}$ is the temperature of the gas. The number density of particles is thus $n_v=N/V$. As the gas particles are assumed non-interacting in our modelling (i.e., rarefied gas), we have assumed the Maxwell–Boltzmann distribution for the velocities. As we will discuss in the following sections, the Maxwell–Boltzmann distribution can be at low temperatures approximated by a narrow Gaussian distribution or in first instance by a Dirac delta distribution. Using the ideal gas law we can then equivalently also investigate the behaviour of the dephasing as a function of the pressure $p=n k_B T_\text{gas}$.  The corresponding averaged dephasing factor $\bar{\Gamma}_{\operatorname{n}}$ can be computed using Eq.~\eqref{eq:GammaDef} by formally replacing $S_{aa}$ with $\bar{S}_{aa}(\omega)$.

The moving external particles can cause a random acceleration noise on the interferometer due to the EM interactions, contributing to a phase fluctuation of the interferometer. 
In the following, we will first explore the behaviour of the non-averaged acceleration PSD  $S_{aa}(b,v,\alpha,\beta;\omega)$ and of the resulting dephasing factor $\Gamma_{\operatorname{n}}$ defined in Eq.~\eqref{eq:GammaDef}. Such an analysis provides an indication of the allowed parameter space of $b,v,\alpha,\beta$ for the dephasing to remain contained (see Secs.~\ref{sec:charged}-\ref{sec:results}). We will then compute the averaged acceleration PSD $\bar{S}_{aa}(\omega)$ and the corresponding averaged dephasing $\bar{\Gamma}_{\operatorname{n}}$ using the probability distributions in Eqs.~\eqref{eq:1}-\eqref{eq:4} for two applications (Sec.~\ref{sec:applications}). 

We suppose the micro-particle and the moving particle can be modelled as either point charges or dipoles and consider two cases for the interactions.\\

\subsection{Charged interferometer}
    
 If the interferometer is charged, the micro-crystal has an overall charge. Then, there can be two EM interactions between the charged interferometer and the external particle:
\begin{enumerate}
    \item A charge-charge interaction between the external charge, $q_\text{ext}$, and the charge of the interferometer, denoted here by $q_\text{int}$. 
    The Coulomb potential gives this interaction
    \begin{equation}\label{eq:Vcc}
        V_\text{cc}(r) = \frac{1}{4\pi\epsilon_{0}} \frac{q_\text{int} q_\text{ext}}{r} \, ,
    \end{equation}
    where both systems are considered as point charges.
    
    \item A charge-dipole interaction between the charge of the interferometer, $q_\text{int}$, and the dipole of the external particle, $\boldsymbol{d}_\text{ext}$.
    This interaction is given by the potential, see~\cite{griffiths2013introduction,jackson_classical_1999}:
    \begin{equation}\label{eq:Vcd}
        V_\text{cd}(r) = \frac{q_\text{int}}{4\pi\epsilon_{0}} \frac{\textbf{d}_\text{ext}\cdot\hat{\textbf{r}}}{r^2} \, .
    \end{equation}
    The external dipole can be either induced or permanent.
\end{enumerate}

\subsection{Neutral interferometer}

 If the interferometer is neutral, there can be two types of EM interactions between the interferometer and the external particle. 
\begin{enumerate}
    \item A dipole-charge interaction between the dipole of the interferometer, $\boldsymbol{d}_\text{int}$ (which can be either permanent or induced by the EM field of the external particle), and the charge of the external particle, $q_\text{ext}$. 
    This interaction is given by the potential:~\cite{griffiths2013introduction}
    \begin{equation}\label{eq:Vdc}
        V_\text{dc}(r) = - \frac{q_\text{ext}}{4\pi\epsilon_{0}} \frac{\textbf{d}_\text{int}\cdot\hat{\textbf{r}}}{r^2} \, ,
    \end{equation}
    where the external particle is assumed to be a point charge.
    \item A dipole-dipole interaction, where the dipole of the interferometer, $\boldsymbol{d}_\text{int}$, and the dipole of the external dipole, $\boldsymbol{d}_\text{ext}$, are assumed to be permanent. This interaction is given by the potential:~\cite{griffiths2013introduction,jackson_classical_1999}
    \begin{align}
        V_\text{dd}(r) = \frac{1}{4\pi\epsilon_{0}} \bigg[ &\frac{\textbf{d}_\text{int}\cdot\textbf{d}_\text{ext}}{r^3} \nonumber \\ &\,\,- \frac{3(\textbf{d}_\text{int}\cdot\hat{\textbf{r}})(\textbf{d}_\text{ext}\cdot\hat{\textbf{r}})}{r^3} \bigg] \, , \label{eq:Vdd}
    \end{align}
    where $\hat{\boldsymbol{r}}$ is the unit vector pointing in the direction from the interferometer centre of mass to the external particle centre of mass.
\end{enumerate}
\subsection{Accelerations due to charged and neutral interactions}

These interaction potentials will cause a random noise on the interferometer $a_\text{noise, EM}$, which will contribute a phase fluctuation that can be detectable by sensing it if we were to use the matter-wave interferometer as a quantum sensor, or this acceleration noise will lead to dephasing the experimental outcome in a QGEM-type experiment for instance.
We now give a brief outline of all the accelerations.
For each of the interactions discussed above,~eqs.~\eqref{eq:Vcc}--\eqref{eq:Vdd}, the corresponding acceleration is found via $\boldsymbol{F}=m\,\boldsymbol{a}$, with $\boldsymbol{F} = -\partial V/\partial\boldsymbol{r}$, where $m$ is the mass of the interferometer. 
\begin{align}
    \boldsymbol{a}_\text{cc} &= \frac{q_\text{int} q_\text{ext}}{4\pi m\epsilon_0} \frac{\hat{\boldsymbol{r}}(t)}{r^2(t)}, \label{eq:a-cc} \\
    \boldsymbol{a}_\text{cd} &= \frac{q_\text{int}}{2 \pi m \epsilon_{0}} \frac{\mathbf{d}_\text{ext} \cdot \hat{\mathbf{r}}(t)}{r^3(t)} \, \hat{\boldsymbol{r}}(t)\, , \label{eq:a-cd} \\
    \boldsymbol{a}_\text{dc} &= \frac{q_\text{ext}}{2 \pi m \epsilon_{0}} \frac{\mathbf{d}_\text{int} \cdot \hat{\mathbf{r}}(t)}{r^3(t)} \, \hat{\boldsymbol{r}}(t)\, , \label{eq:a-dc} \\
    \boldsymbol{a}_\text{dd} 
    &= \frac{3 \,\hat{\boldsymbol{r}}}{4\pi\epsilon_0 m} \left[ \frac{\boldsymbol{d}_\text{ext} \cdot \boldsymbol{d}_\text{int}}{r^4(t)} - 3 \frac{ (\boldsymbol{d}_\text{ext} \cdot \hat{\boldsymbol{r}}) (\boldsymbol{d}_\text{int}\cdot \hat{\boldsymbol{r}})}{r^4(t)} \right] \, . \label{eq:a-dd}
\end{align}
The vector $\hat{\boldsymbol{r}}$ is time-dependent and given by $\hat{\boldsymbol{r}}(t) = \boldsymbol{r}(t)/r(t)$, with $r(t)=\abs{\boldsymbol{r}(t)}$ (as defined in eq.~\eqref{eq:distancevector}).
The dipole can be either permanent or induced, which we distinguish with the notation $\boldsymbol d(i)$ and $\boldsymbol d(p)$, respectively.

Since the interferometer is created in one dimension, e.g. $\hat{x}$-direction in our case, the only relevant acceleration is in the $\hat{x}$-direction, and we are specifically interested in $a_x(t)$.
The acceleration is found separately now for every interaction in Eqs.~\eqref{eq:Vcc}--\eqref{eq:Vdd}, and some assumptions are made on the orientation of the dipole vectors for simplification and to estimate the upper bound on the dephasing.

The mean value of the noise can be assumed to be zero, e.g. 
\begin{equation}
\mathbb{E}[a_\text{noise, EM}]=\mathbb{E}[\delta \phi]=0\,.
\end{equation}
This is because the zero point, or the baseline of the phase, can be calibrated using an axillary experiment, so the contribution of the mean value of every noise will be considered in the offset of the baseline in our current analysis.
Therefore, we will here focus on finding the autocorrelation of the acceleration noise at different times, $\mathbb{E}[a_\text{noise, EM}(t_1),a_\text{noise, EM}(t_2)]$, see eq.~\eqref{eq:PSD_wienerkhinchin}.
Taking the Fourier transform to find $a_x(\omega)$ on the time domain $[0,T]$, we can use eq.~\eqref{eq:saa_def} to find the noise PSD $S(\omega)=\abs{a_x(\omega)}^2/T$.
Combining the noise PSD with the transfer function gives the dephasing using eq.~\eqref{eq:GammaDef}.
The results for the noise are given in Sec.~\ref{sec:charged},~\ref{sec:neutral}, and the detailed calculations are presented in Appendix~\ref{app:detail_calc}.


\section{Dephasing of a Charged interferometer}\label{sec:charged}

If there is some charge on the interferometric particle, then charge-charge interactions and charge-external dipole interactions will generate acceleration noise in the system. We consider the charge-charge interaction and the charge-dipole interaction between the interferometer and the external particle, where we investigate separately the dipole of the external particle to be permanent or induced
\footnote{These interactions depend on the relative sign of the charges and dipoles (whether they are attractive or repulsive). 
In this section, we do not specify the relative sign as we are interested in finding the dephasing, and we note that this is proportional to the acceleration \textit{squared}.
The dephasing is therefore independent of the sign of the interaction potential.}.


\subsection{Dephasing due to internal charge and external charge interaction}\label{subsec:charge-charge}

First of all, we will assume the simplest case of a charged matter-wave interferometer: the matter-wave interferometer has a charge $q_\text{int}$, which is interacting with ambient charge $q_\text{ext}$ that is moving with a constant velocity and has the closest approach $b$ from the interferometer's centre of mass.
The acceleration due to the charge-charge interaction in the $\hat{x}$-direction is given by:
\begin{equation}\label{eq:ax-cc}
    a_x^\text{cc}(t) = \frac{q_\text{int}q_\text{ext}}{4\pi \epsilon_0 m b^2} \frac{\cos(\alpha) + (vt/b) \cos(\beta)}{(1+(vt/b)^2)^{3/2}}.
\end{equation}
Finding $S_{aa}$ from $a_x(\omega)$ and putting it in Eq.~\eqref{eq:GammaDef} yields the dephasing:
\begin{equation}\label{eq:gamma-cc}
\begin{aligned}
        & \Gamma_{\operatorname{n}}^\text{cc} = \frac{q_\text{int}^2q_\text{ext}^2(\Delta x)^2}{\pi^3 \hbar^2 \epsilon_0^2 T v^4 t_a^4} \int_{\omega_{\operatorname{min}}}^{\infty}\omega^{-4} \operatorname{sin}^2\left(\frac{\omega}{2}(2t_a+t_e)\right) \\ 
        &{\operatorname{sin}}^4\left(\frac{t_a \omega}{2}\right)
        \left[\cos^2(\alpha)K_{1}^2\left(\frac{b\omega}{v}\right)+\cos^2(\beta)K_{0}^2\left(\frac{b\omega}{v}\right)\right]d\omega.
\end{aligned}
\end{equation}
with $K_\nu$ is the modified Bessel function (see appendix~\ref{app:detail_calc}).
The dephasing depends on the charges $q_\text{int}$ and $q_\text{ext}$, the size of the spatial superposition and the time during which it is created, the impact parameter and the velocity of the external particle.
Furthermore, the time $T$, the time over which the phase fluctuations are averaged, and the minimum frequency $\omega_\text{min}$ are determined by the experimental setup, impact parameter $b$ and the velocity $v$ of the external particle.
The integral in eq.~\eqref{eq:gamma-cc} is solved numerically, and the resulting dephasing is plotted in Sec.~\ref{sec:results} (Fig. \ref{fig:results-charged-int}(a)) for a specific interferometer scheme.

\subsection{Dephasing due to internal charge and external dipole interaction}\label{section:Dipole_Charged Test-Mass}

This section will consider a charged interferometer particle interacting with an external dipole. 
The dipole could arise from any ambient gas particle inside the vacuum chamber. We will only consider the acceleration noise due to one such external particle.
The water vapour left in the vacuum chamber consists of the water molecules that carry a permanent dipole moment.
Left-over air molecules in the vacuum chamber, such as dinitrogen, carbon dioxide, argon and dioxygen, are polarisable 
 and could thus have an induced dipole moment from the charge of the interferometer.
The acceleration due to the charge-dipole interaction is given in Eq.~\eqref{eq:a-cd}, and the acceleration depends on whether the dipole is induced or permanent. 

\subsubsection{Permanent external dipole}

Suppose the external particle has a permanent dipole, such as in the case of water vapour. In that case, the dipole moment magnitude can be taken to be the experimentally determined value, e.g. $\boldsymbol{d}_\text{ext}^\text{(p)} = 6.19 \times 10^{-30} \, \si{\coulomb\metre}$ in the case of water vapour (the superscript $p$ standing for permanent)~\cite{lide2004crc}.
Taking the worst-case scenario, we assume that at the point of closest approach $b$, the dipole vector of the external particle is aligned with the vector $r(t)$.

Assuming that the particle moves very slowly, we maximize the acceleration by taking the dipole vector and the vector $\boldsymbol{r}(t)$ to align during the experimental time $\tau= t_f- t_i$. 
The acceleration in the $x$-direction can be found to be:
\begin{equation}\label{eq:ax-cd(per)}
    a_x^\text{cd(p)}(t) = \frac{q\abs{\mathbf{d}_\text{ext}}}{2 \pi m b^3 \epsilon_{0}} \frac{\cos(\alpha)+(vt/b)\cos(\beta)}{(1+(vt/b)^2)^2}\,.
\end{equation}
Its Fourier transforms and the resulting PSD of the noise are given in Appendix~\ref{app:detail_calc}.
The dephasing  is then given by Eq.~\eqref{eq:GammaDef}, the full integral will be:
\begin{equation}
\begin{aligned}\label{eq:gamma-cd(per)}
    \Gamma_{\operatorname{n}}^\text{cd(p)} = & \frac{q^2 |\mathbf{d}_{\operatorname{ext}}|^2 (\Delta x)^2}{2\pi^2 \hbar^2\epsilon_{0}^2 T v^5 b t_a^4} \int_{\omega_{min}}^{\infty}\omega^{-3} \operatorname{sin} ^2\left(\frac{\omega}{2}(2t_a+t_e)\right)\\
    & \operatorname{sin}^4\left(\frac{t_a \omega}{2}\right) \Biggr[\cos^2(\alpha) K_{3/2}^2\left(\frac{b\omega}{v}\right) \\
    & \qq{}\qq{} \qq{} \qq{}+ \cos^2(\beta) K_{1/2}^2\left(\frac{b\omega}{v}\right) \Biggr]d\omega \, ,
\end{aligned}
\end{equation}
where $K_\nu$ is the modified Bessel function (see appendix~\ref{app:detail_calc}).
The results of the integration with the parameters of a specific interferometer setup are given in Sec.~\ref{sec:results} (Fig. \ref{fig:results-charged-int}(b)).

\subsubsection{Induced external dipole}

Left-over air molecules in the vacuum chamber have a polarisability of $\alpha_\text{pol} \sim 10^{-40}\,\si{\ampere\squared\second^4\per\kilogram}$~\footnote{
Generally this polarizability is expressed in CGS units,  $\alpha_\text{pol}' = 1.5-2.5$ {\AA}$^3$~\cite{atmosphere,pollistx}, where 1 {\AA}$^3 = 10^{-24} \,\si{\centi\metre\cubed}$.
Which is related to the polarizability in SI units, $\si{\ampere\squared\second^4\per\kilogram}$, via $\alpha_\text{pol} = 4\pi\epsilon_0 \alpha_\text{pol}'$
The particle polarizability is related to the average electric susceptibility of a medium, $\epsilon$, via the Clausius–Mossotti relation, see eq.~\eqref{Classius-Mossotti}.
}.
If the dipole moment in the external particle with polarisability is induced due to the electric field $\boldsymbol{E}$ from the charged interferometer, then the result is slightly different.
The induced dipole by a charge $q_\text{int}$ is given by:
\begin{equation}
    \boldsymbol{d}_\text{ext}^\text{(i)} = \alpha_\text{pol} \boldsymbol{E} = \alpha_\text{pol} \frac{q_\text{int}}{4\pi\epsilon_0 r^2} \hat{\boldsymbol{r}} \,,
\end{equation}
where the superscript (i) stands for induced.
The acceleration in the $\hat{x}$-direction thus becomes:
\begin{align}
        a_x^\text{cd(i)}(t) &= \frac{q_\text{int}^2\alpha_\text{pol}}{8 \pi^2 m b^5 \epsilon_{0}^2} \frac{\cos(\alpha)+(vt/b)\cos(\beta)}{(1+(vt/b)^2)^2} \frac{1}{r^2(t)}, \nonumber \\
        &= \frac{q_\text{int}^2\alpha_\text{pol}}{8 \pi^2 m b^5 \epsilon_{0}^2} \frac{\cos(\alpha)+(vt/b)\cos(\beta)}{(1+(vt/b)^2)^3} .\label{eq:ax-cd(ind)}
\end{align}
Appendix~\ref{app:detail_calc} presents the Fourier transform and PSD.
The dephasing is then found to be:
\begin{equation}
    \begin{aligned}\label{eq:gamma-cd(ind)}
    \Gamma_{\operatorname{n}}^\text{cd(i)} =  \frac{q^4\alpha^2_\text{pol} (\Delta x)^2}{512 \pi^4 \hbar^2 \epsilon_{0}^4  v^7 b^3 T t_a^4} &\int_{\omega_{\operatorname{min}}}^{\infty} d \omega \frac{1}{\omega} \sin ^4\left(\frac{t_a \omega}{2}\right) \\ 
    \sin ^2\left(\frac{\omega}{2}(2t_a+t_e)\right)&\Biggr[\cos^2(\alpha)K^2_{5/2}\left(\frac{b\omega}{v}\right) \\
    & + \cos^2(\beta)K^2_{3/2}\left(\frac{b\omega}{v}\right)\Biggr].
    \end{aligned}
\end{equation}
$K_\nu$ is the modified Bessel function (see appendix~\ref{app:detail_calc}).
Again, this expression can be solved numerically to obtain the dephasing. 
The results of the integration with the parameters of a specific interferometer setup are given in Sec.~\ref{sec:results} (figure \ref{fig:results-charged-int}(c)).

Eq.~\eqref{eq:gamma-dc(ind)} differs from the permanent dipole result in Eq.~\eqref{eq:gamma-dc(per)} since the induced dipole is assumed to be caused by the interferometer charge. As a result, Eq.~\eqref{eq:gamma-dc(ind)} depends on the interferometer charge and the polarisability of the external particle, $\alpha_\text{pol}$. The magnitude of the induced dipole is distance-dependent, resulting in a different $v,b$-dependence of Eq.~\eqref{eq:gamma-dc(ind)} compared to the permanent dipole dephasing result.


\section{Dephasing of a Neutral interferometer}\label{sec:neutral}

In this section, we will assume that the micro-particles in the interferometer have a dipole $d_\text{int}$ (either permanent or induced) and that it is interacting with an external charged particle (of charge $q_\text{ext}$) or external dipole ($\boldsymbol{d}_\text{ext}$).
We consider separately the microcrystal's permanent and induced dipole moments in the presence of a charge.

Although the dephasing results obtained in this section are generic, we will apply them to the QGEM experiment where neutral diamond-type crystals of mass $\sim10^{-15}\,\si{\kilo\gram}$ are considered.
We briefly discuss the two different dipoles of the micro-particle in this context.

\begin{itemize}
    \item{Permanent dipole}: The microcrystal may have a permanent dipole due to the surface's impurities or volume.
The dipole moment of silica-type material, ($\text{SiO}_{2}$), was experimentally measured in \cite{Afek:2021bua}, which
showed that the material exhibits no clear correlation between mass and dipole moment. 
That being said, the micro-particle spheres of size $10\,\si{\micro\metre}$, $15\,\si{\micro\metre}$ and $20\,\si{\micro\metre}$ had permanent dipoles varying from $\sim 1000-5000\,e\,\si{\micro\metre}$, $\sim 200-8000\,e\,\si{\micro\metre}$ and $\sim 8500\pm1500 \,e\,\si{\micro\metre}$ (for simplicity we use $e\,\si{\micro\metre}$ to express the dipole, with $e$ a single electric charge such that $1\,e \,\si{\micro\metre}\approx 1.6 \times 10^{-25} \,\si{\coulomb\metre}$). 
This shows at least an order of magnitude uncertainty in the permanent dipole magnitude. 
The study showed that this material's permanent dipole moments exhibit a volume scaling, leaving the question of whether the permanent dipole moment scales with the volume.
For the test-mass of radius $r=0.5\,\si{\micro\metre}$ (corresponding to $\sim10^{-15}\,\si{\kilo\gram}$ for a spherical diamond test mass), we take the dipole moment to be $|\textbf{d}_\text{int}^{(p)}|\sim 0.1 \,e\,\si{\micro\metre}$, thus assuming a volume scaling, and using the experimental data in \cite{Afek:2021bua} as a benchmark. 

    \item{Induced dipole}: The diamond-type crystal is a dielectric material, and the crystal has a polarizability $\alpha$ (in SI-units $\si{\ampere\squared\second^4\per\kilogram}$).
In particular, in isotropic media, a local electric field will produce a local dipole in each atom of the crystal's lattice~\cite{jackson_classical_1999}:
\begin{equation}
    \boldsymbol{d}_\text{int}^{(i)}=N \alpha_\text{pol} \, \boldsymbol{E}_{\text{loc}}.
    \label{eq:induced_63}
\end{equation}
The local polarizability (which we denote $\alpha_\text{pol}$) of the atoms is related to the polarisability of the medium via the Classius-Mossotti relation:
\begin{equation}
    \frac{n \, \alpha_\text{pol}}{3 \epsilon_{0}}=\frac{\epsilon_{r}-1}{\epsilon_{r}+2} \,,
    \label{Classius-Mossotti}
\end{equation}
with $n$ the number density of atoms, $n=3N/4\pi R^3$ for spherical masses ($N$ the number of atoms, $R$ the radius).
As a result, the dipole for a spherical diamond crystal due to an external point charge $q_\text{ext}$ is given by:
\begin{equation}
    \boldsymbol{d}_\text{int}^{(i)} = \frac{\epsilon_r-1}{\epsilon_r+2} \frac{q_\text{ext} R^3}{r^2} \hat{\boldsymbol{r}} \, ,
\end{equation}
with $\epsilon_r$ relative permittivity of the medium. 
\end{itemize}

\subsection{Dephasing due to internal dipole and charge interaction}\label{subsec:dipole-charge}

We now consider the dephasing due to an internal dipole interacting with an external charge.
The acceleration was given in Eq.~\eqref{eq:a-dc}, the $x$-component of which depends on whether the dipole is permanent or induced.

\subsubsection{Permanent  dipole of a micro-particle}

Suppose the interferometer's micro-particle has a permanent dipole moment, similar to that of silica-type material~\cite{Afek:2021bua}. In that case, the dipole moment magnitude can be the experimentally determined value, e.g. $\sim e \, \si{\micro\metre}$ for micro-spheres, assuming a volume scaling as discussed previously.
Furthermore, if we assume complete control of the interferometer particle, we can take the direction of the intrinsic dipole moment to be aligned with the $\hat{z}$-axis.
As detailed in Appendix~\ref{app:detail_calc}, 
in this scenario the component of the acceleration in the $\hat{x}$-direction is:
\begin{equation}\label{eq:ax-dc(per)}
\begin{aligned}
        a_x^\text{d(p)c}(t)
        = \frac{q_\text{ext}|\mathbf{d}_\text{int}|}{2 \pi m b^3 \epsilon_{0}} &\frac{\cos(\alpha)+(vt/b)\cos(\beta)}{(1+(vt/b)^2)^{5/2}} \\
    &\bigg[ \cos(\theta_0) + (vt/b)\cos(\gamma)\bigg],
\end{aligned}
\end{equation}
where $\gamma$ is the angle between the velocity vector and the $\hat{z}$-axis and $\theta_0$ is the angle between the vector $b$ and the $\hat{z}$-axis (the angles $\gamma$, $\theta_0$ are similar to the projection angles $\alpha$, $\beta$, respectively, in figure~\ref{fig:external particle}, but $\gamma$, $\theta_0$ are projection angles on the $\hat{z}$-axis rather than on the $\hat{x}$-axis, they are discussed in more detail in Appendix~\ref{app:detail_calc}). 

The Fourier transform $a(\omega)$ gives an expression for the PSD, $S_{aa}(\omega)$, and from these expressions we find the dephasing parameter, see Appendix~\ref{app:detail_calc}.
Taking the worst-case dephasing based on the angles $\alpha,\beta,\gamma,\theta_0$ (see Appendix~\ref{app:proj_angles}, where we optimise these angles to see the largest dephasing), the dephasing is given as:
\begin{equation}
\begin{aligned}\label{eq:gamma-dc(per)}
    \Gamma_{\operatorname{n}}^\text{d(p)c} = & 
    \frac{q_\text{ext}^2|\mathbf{d}_\text{int}|^2 (\Delta x)^2}{9\pi^3\epsilon_0^2 \hbar^2 v^6 T t_a^4}
    \int^\infty_{\omega_\text{min}} \frac{1}{\omega^2} \sin^4\left(\frac{t_a \omega}{2}\right) \\
    & \Bigg[ 
    K_{2}^2\left(\frac{b\omega}{v}\right) + \left( 2 + \frac{b^2\omega^2}{v^2} \right) K_{1}^2\left(\frac{b\omega}{v}\right) \\
    &+ K_{0}^2\left(\frac{b\omega}{v}\right) - 2 \frac{b\omega}{v} K_{0}\left(\frac{b\omega}{v}\right) K_{1}\left(\frac{b\omega}{v}\right) 
    \Bigg] \\
    &\qq{}\sin^2\left(\frac{\omega}{2}(2t_a+t_e)\right)
    \dd{\omega}
\end{aligned}
\end{equation}
which can be solved numerically and where $K_\nu$ is the modified Bessel function (see Appendix~\ref{app:detail_calc}).
The results of the integration with the parameters of a specific interferometer setup are given in Sec.~\ref{sec:results} (figure \ref{fig:results-neutral-int}(a)).

\subsubsection{Induced dipole of a micro-particle}

If the dipole is induced by the electric field of an external charge $q_\text{ext}$, its magnitude is given by~\cite{jackson_classical_1999,griffiths2013introduction}:
\begin{equation}
    \boldsymbol{d}_\text{int}^\text{(i)} = 
    \frac{\epsilon_r-1}{\epsilon_r+2} \frac{q_\text{ext} R^3}{r^2(t)} \hat{\boldsymbol{r}} \, ,
\end{equation}
with the relative permittivity $\epsilon_r\sim5.7$ for diamond and $R$ the radius of the spherical diamond (the superscript (i) indicates the induced dipole).
The acceleration in the $\hat{x}$-direction then becomes:
\begin{equation}\label{eq:ax-dc(ind)}
\begin{aligned}
        a_x^\text{d(i)c}(t)
        &= \frac{\epsilon_r-1}{\epsilon_r+2} \frac{q_\text{ext}^2 R^3}{2 \pi \epsilon_{0} m b^5 } \frac{\cos(\alpha)+(vt/b)\cos(\beta)}{(1+(vt/b)^2)^3} \, .
\end{aligned}
\end{equation}
The dephasing parameter is derived in detail in Appendix~\ref{app:detail_calc}, and is given by:
\begin{equation}
    \begin{aligned}\label{eq:gamma-dc(ind)}
    \Gamma_{\operatorname{n}}^\text{d(i)c} = 
        &\left(\frac{\epsilon_r-1}{\epsilon_r+2}\right)^2 \frac{q_\text{ext}^4 R^6 (\Delta x)^2}{32 \pi^2 \epsilon_{0}^2 \hbar^2 v^7 b^3 T t_a^4} \int_{\omega_\text{min}}^\infty \frac{1}{\omega} \\ & \Biggr[\cos^2(\alpha)K^2_{5/2}\left(\frac{b\omega}{v}\right) + \cos^2(\beta)K^2_{3/2}\left(\frac{b\omega}{v}\right)\Biggr] \\&\qq{}\sin ^4\left(\frac{t_a \omega}{2}\right) \sin ^2\left(\frac{\omega}{2}(2t_a+t_e)\right) \dd{\omega},
    \end{aligned}
\end{equation}
and $K_\nu$ is the modified Bessel function (see appendix~\ref{app:detail_calc}).
Again, this expression can be solved numerically to obtain the dephasing. 
The results of the integration with the parameters of a specific interferometer setup are given in Sec.~\ref{sec:results} (Fig. \ref{fig:results-neutral-int}(b)).
Eq.~\eqref{eq:gamma-dc(ind)} differs from the permanent dipole result in Eq.~\eqref{eq:gamma-dc(per)} since the induced dipole is assumed to be induced by the external charge. As a result, Eq.~\eqref{eq:gamma-dc(ind)} depends on the external charge and the relative permittivity and radius of the interferometer. 
The magnitude of the induced dipole is distance-dependent, resulting in a different $v,b$-dependence of Eq.~\eqref{eq:gamma-dc(ind)} compared to the permanent dipole dephasing result of Eq.~\eqref{eq:gamma-dc(per)}.


\subsection{Dephasing due to internal dipole and external dipole}\label{subsec:dipole-dipole}

The acceleration from the dipole-dipole interaction was given in Eq.~\eqref{eq:a-dd}.
For large $r$, we can see that the other interactions dominate this interaction since it goes as $1/r^4$.
Sticking with the assumptions made in previous sections, that $\boldsymbol{d}_\text{ext}$ aligns with the $\hat{z}$-direction, and that for a short interaction time $\tau$, $\boldsymbol{d}_\text{int}$ approximately aligns with $\hat{\boldsymbol{r}}(t)$, the acceleration in the $\hat{x}$-direction is:
\begin{align}
    \boldsymbol{a}_x^\text{dd}(t)
    = \frac{6 \abs{\boldsymbol{d}_\text{ext}} \abs{ \boldsymbol{d}_\text{int}}\cos(\theta_0)}{4\pi\epsilon_0 m b^4} &\frac{\cos(\alpha) + (vt/b) \cos(\beta)}{(1+(vt/b)^2)^{5/2}} \, . \label{eq:ax-dd}
\end{align}
More details are given in Appendix~\ref{app:detail_calc}, where the Fourier transform of the acceleration and the resulting PSD is given. 
From the PSD and the transfer function, the dephasing is found to be:
\begin{equation}
    \begin{aligned}\label{eq:gamma-dd}
    \Gamma_{\operatorname{n}}^\text{dd} = &
    \frac{4 \abs{\boldsymbol{d}_\text{ext}}^2 \abs{\boldsymbol{d}_\text{int}}^2 (\Delta x)^2 \cos^2(\theta_0)}{\pi^3 \hbar^2 \epsilon_0^2 b^2 v^6 T t_a^4}
    \int_{\omega_\text{min}}^\infty \frac{1}{\omega^2} 
    \\ &
    \Bigg[ \cos^2(\alpha) K_2^2\left(\frac{b \omega}{v}\right) + \cos^2(\beta) K_1^2\left(\frac{b \omega}{v}\right) \Bigg] 
    \\ &
    \sin ^4\left(\frac{t_a \omega}{2}\right) \sin ^2\left(\frac{\omega}{2}(2t_a+t_e)\right) \dd{\omega} ,
    \end{aligned}
\end{equation}
see Appendix~\ref{app:detail_calc} for the detailed derivation. 
However, the result is an approximation given the assumptions made on this setup's dipole moments and velocities.
Ref.~\cite{Fragolino:2023agd} provides a more general expression of the decoherence due to dipole-dipole interactions for spatial interferometers.
The results of the integration with the parameters of a specific interferometer setup are given in Sec.~\ref{sec:results} (figure \ref{fig:results-neutral-int}(c)). 

\section{Dephasing Results}\label{sec:results}

\begin{figure*}[t]
\centering
    \includegraphics[width=0.47\linewidth]{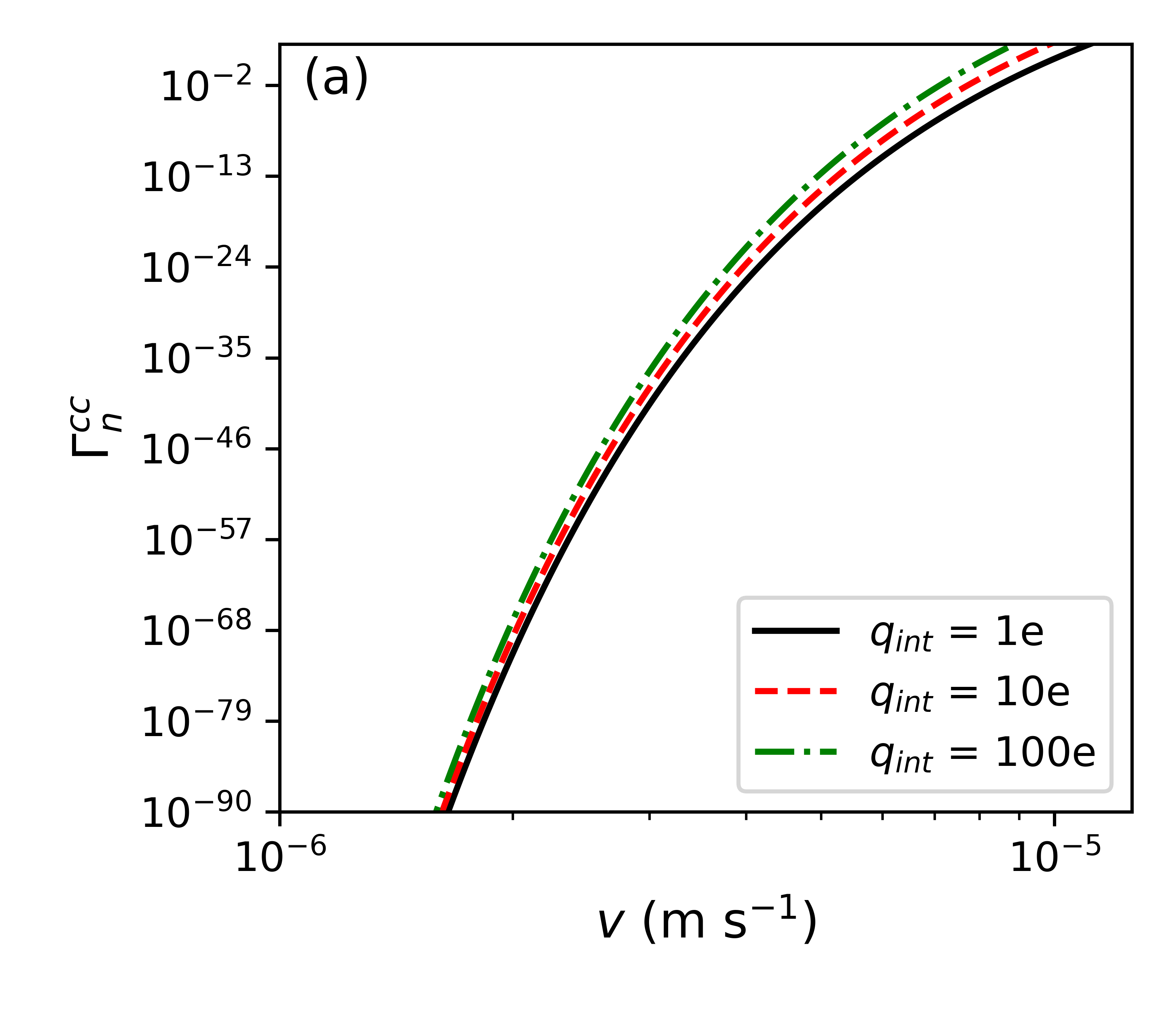}\hfil
    \includegraphics[width=0.47\linewidth]{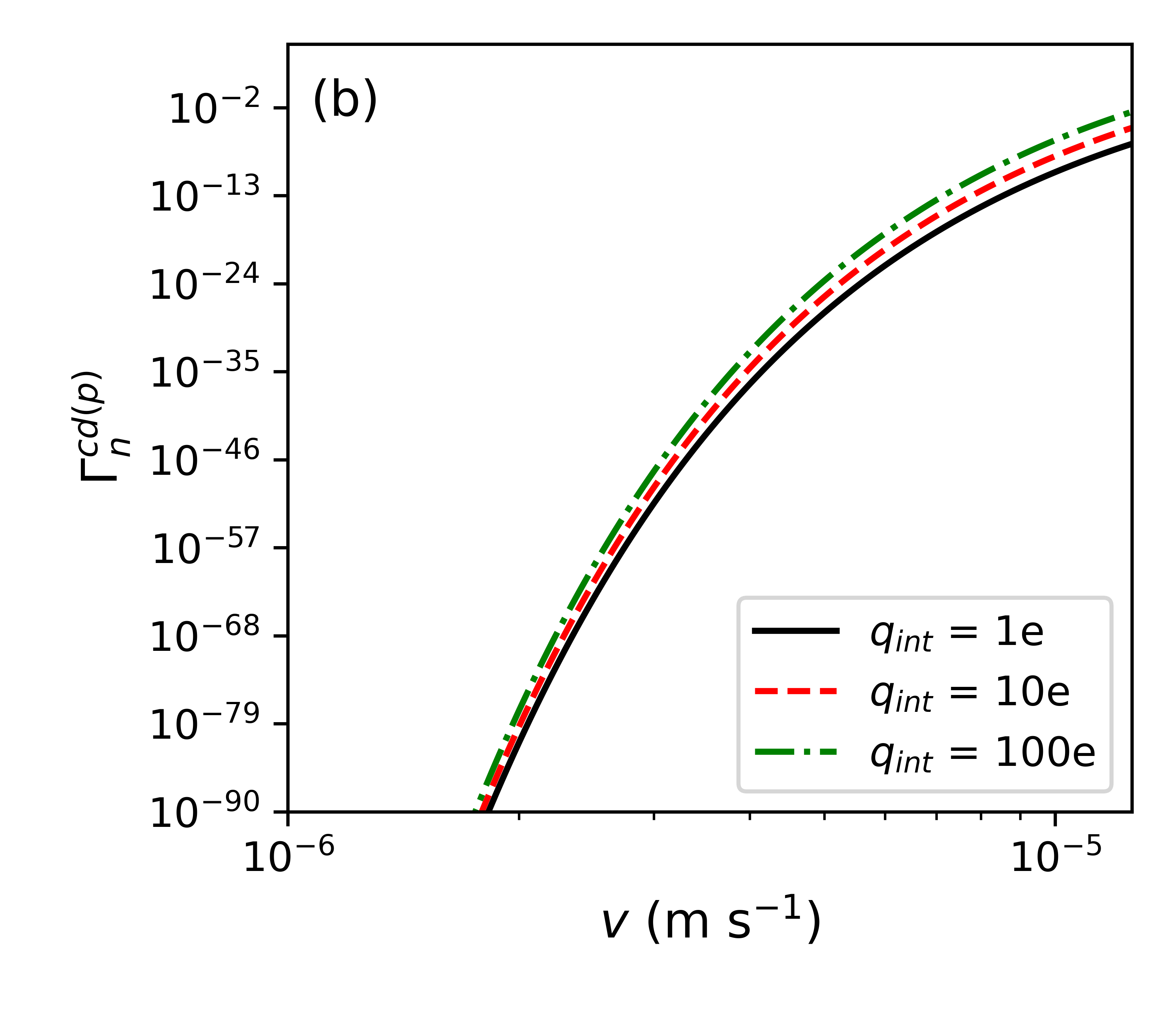}\par\medskip
    \includegraphics[width=0.47\linewidth]{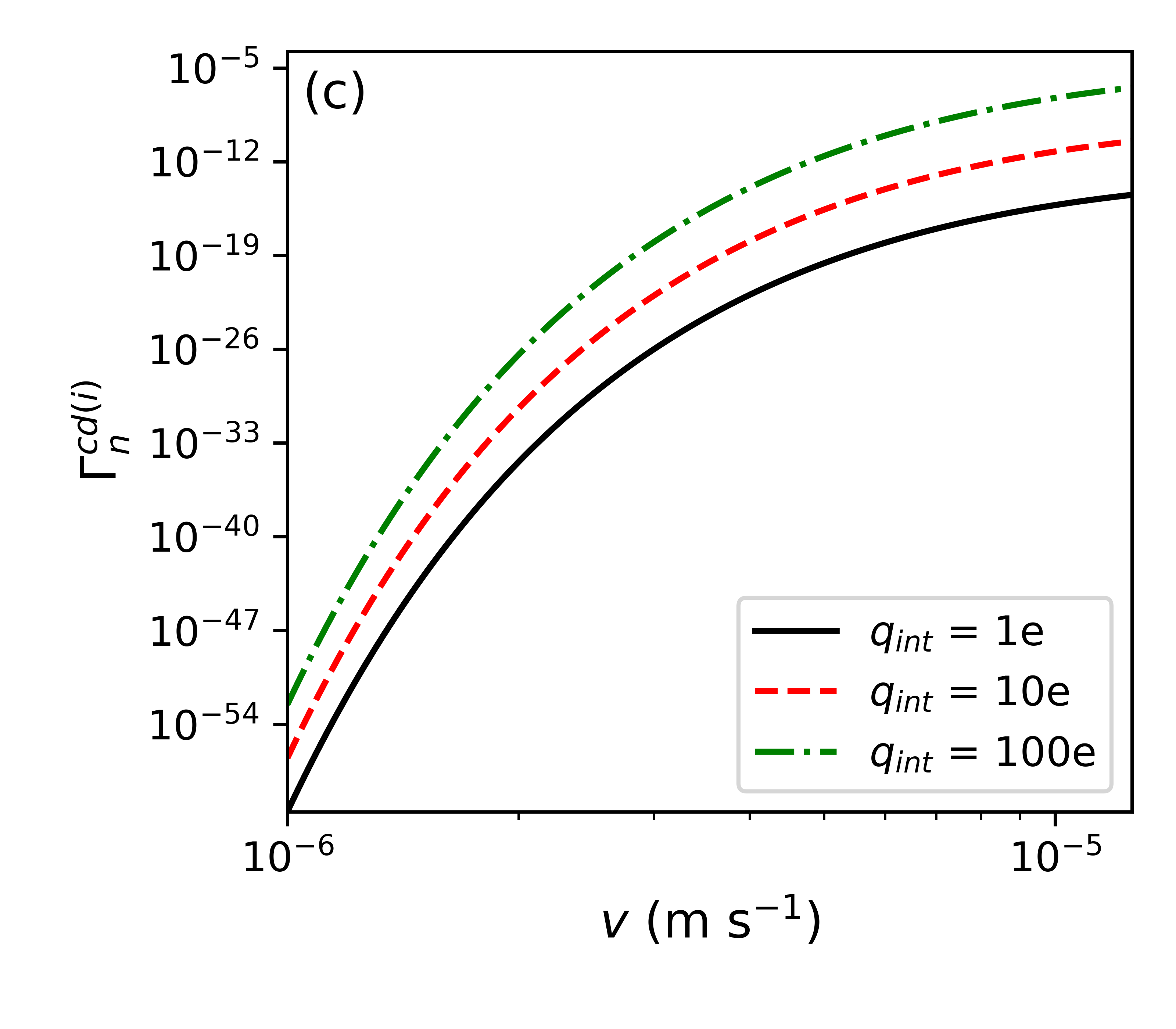}\hfil
    \includegraphics[width=0.47\linewidth]{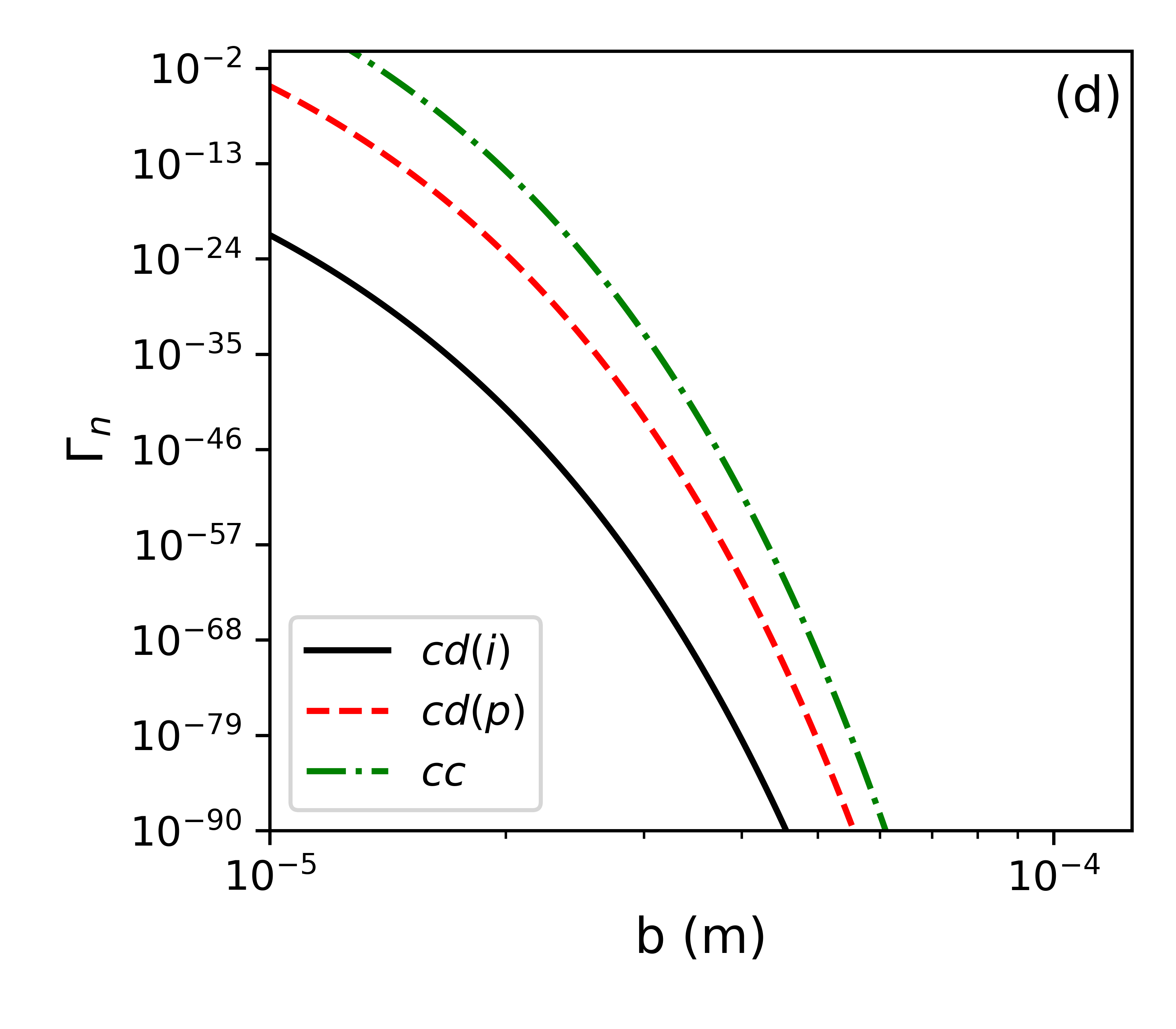}
\caption{Dephasing as a function of the velocity of the external particle due to (a) the charge-charge interaction (cc), (b) the charge-permanent dipole interaction (cd(p)) and (c) the charge-induced dipole interaction (cd(i)), for a different number of interferometer charges.
The expressions summarised in Tab.~\ref{table:results} are plotted, where the integral is solved numerically.
With projection angles $\alpha=\beta=0$ and $T = b/v$. We have taken $\Delta x=20{\rm \mu m}$ in all the plots.
In Fig.(a) $b=100\,\si{\micro\metre}$ and the external charge is that of an electron, in Fig.(b) $b=100\,\si{\micro\metre}$ and the external dipole moment $d_\text{ext}$ is that of a water molecule, in Fig.(c) $b=30\,\si{\micro\metre}$ and the external particle's polarizability is that of the dinitrogen.
Fig.(d) shows the different dephasing for a charged interferometer as a function of the impact parameter, for $v=1\,\si{\micro\metre\per\second}$, $q_\text{int}=e$ and the interaction-specific parameters discussed above. 
This plot (d) shows that the interferometer could detect a $q_\text{ext}=e$ charged particle moving with a velocity $v=1 \,\si{\micro\metre\per\s}$ in the vicinity of $10\,\si{\micro\metre}$. The Coulomb interaction provides the largest dephasing compared to the charge dipole interactions.
In all the plots we have ensured $T\sim b/v > \tau\sim 3\,\si{\s}$.}
\label{fig:results-charged-int}
\end{figure*}

The dephasings found in this paper are summarised in Tab.~\ref{table:results}.
The dephasing expressions are plotted in Figs.~\ref{fig:results-charged-int} and~\ref{fig:results-neutral-int}, which show that the dephasing increases for increasing velocity and decreasing impact parameter.
The dephasing expressions show a solid inverse velocity dependence, but there is also a velocity dependence inside the integral part, which has been solved numerically to obtain the figures.
Therefore, from the results in the table, it is tricky to draw any conclusions about physics. 


\begin{table*}[t] 
    \centering
    \setlength{\tabcolsep}{12pt} 
    \renewcommand{\arraystretch}{2}
    \begin{tabular}{p{0.16\linewidth} |c}
    \hline
    \hline
    Interaction \newline Notation & Dephasing Expression \\
    \hline 
    \hline
    charge -- charge \newline cc & $\Gamma_{\operatorname{n}}^\text{cc} = \frac{q_\text{int}^2 q_\text{ext}^2 (\Delta x)^2}{\pi^3 \hbar^2 \epsilon_0^2 T v^4 t_a^4} \int_{\omega_{\operatorname{min}}}^{\infty} \omega^{-4}
        \left[\cos^2(\alpha)K_{1}^2\left(\frac{b\omega}{v}\right)+\cos^2(\beta)K_{0}^2\left(\frac{b\omega}{v}\right)\right] f(\omega)d\omega.$ \\
    charge -- \newline permanent dipole \newline cd(p) & $\Gamma_{\operatorname{n}}^\text{cd(p)} = \frac{q^2 |\mathbf{d}_{\operatorname{ext}}|^2 (\Delta x)^2}{2\pi^2 \hbar^2\epsilon_{0}^2 T v^5 b t_a^4} \int_{\omega_{min}}^{\infty}\omega^{-3} \left[\cos^2(\alpha) K_{3/2}^2\left(\frac{b\omega}{v}\right)+ \cos^2(\beta) K_{1/2}^2\left(\frac{b\omega}{v}\right) \right] f(\omega) d\omega$ \\
    charge -- \newline induced dipole \newline cd(i) & $\Gamma_{\operatorname{n}}^\text{cd(i)} =  \frac{q^4\alpha^2_\text{pol} (\Delta x)^2}{512 \pi^4 \hbar^2 \epsilon_{0}^4  v^7 b^3 T t_a^4} \int_{\omega_{\operatorname{min}}}^{\infty} \omega^{-1} \left[\cos^2(\alpha)K^2_{5/2}\left(\frac{b\omega}{v}\right) + \cos^2(\beta)K^2_{3/2}\left(\frac{b\omega}{v}\right)\right] f(\omega) \dd{\omega}$ \\
    \hline
    permanent dipole -- \newline charge \newline  d(p)c & 
    $\!\begin{aligned}[t] \Gamma_{\operatorname{n}}^\text{d(p)c} &=
    \frac{4 q_\text{ext}^2|\mathbf{d}_\text{int}|^2 (\Delta x)^2}{9\pi^3\epsilon_0^2 \hbar^2 v^6 t_a^4}
    \int^\infty_{\omega_\text{min}} \omega^{-2} \Big[
    \cos^2(\alpha) 
    \Big( K_{2}^2\left(\frac{b\omega}{v}\right) \cos^2(\theta_0) + K_{1}^2\left(\frac{b\omega}{v}\right) \cos^2(\gamma) \Big) \\ &+ \cos^2(\beta) 
    \Big( K_{1}^2\left(\frac{b\omega}{v}\right) \cos^2(\theta_0) + \Big[\frac{b\omega}{v} K_1\left(\frac{b\omega}{v}\right) - K_0\left(\frac{b\omega}{v}\right)\Big]^2 \cos^2(\gamma) \Big)
    \Big] f(\omega) \dd{\omega}\end{aligned}$ \\
    induced dipole -- \newline charge \newline d(i)c  & $\Gamma_{\operatorname{n}}^\text{d(i)c} = 
    \left[\frac{\epsilon_r-1}{\epsilon_r+2}\right]^2 \frac{q_\text{ext}^4 R^6 (\Delta x)^2}{32 \pi^2 \epsilon_{0}^2 \hbar^2 v^7 b^3 T t_a^4} \int_{\omega_\text{min}}^\infty \omega^{-1} \left[\cos^2(\alpha)K^2_{5/2}\left(\frac{b\omega}{v}\right) + \cos^2(\beta)K^2_{3/2}\left(\frac{b\omega}{v}\right)\right] f(\omega) \dd{\omega} $ \\
    dipole -- dipole \newline dd & $\Gamma_{\operatorname{n}}^\text{dd} =
    \frac{4 \abs{\boldsymbol{d}_\text{ext}}^2 \abs{\boldsymbol{d}_\text{int}}^2 (\Delta x)^2 \cos^2(\theta_0)}{\pi^3 \hbar^2 \epsilon_0^2 b^2 v^6 T t_a^4}
    \int_{\omega_\text{min}}^\infty \omega^{-2}
    \left[ \cos^2(\alpha) K_2^2\left(\frac{b \omega}{v}\right) + \cos^2(\beta) K_1^2\left(\frac{b \omega}{v}\right) \right] f(\omega) \dd{\omega}$ \\
    \hline
    \hline
    \end{tabular}
    \caption{Summary of the dephasing equations, the results were presented in (from top to bottom) Eqs.~\eqref{eq:gamma-cc},~\eqref{eq:gamma-cd(per)},~\eqref{eq:gamma-cd(ind)} for the charged interferometer and Eqs.~\eqref{eq:gamma-dc(per)},~\eqref{eq:gamma-dc(ind)} and~\eqref{eq:gamma-dd} for the neutral interferometer. 
    $f(\omega)$ denotes the sinusoidal part of the transfer function, i.e., $f(\omega) \equiv \sin^4(\frac{t_a \omega}{2}) \sin ^2(\frac{\omega}{2}[2t_a+t_e])$, $\alpha$ and $\beta$ are the projection angles of the impact parameter vector and velocity on the $\hat{x}$-axis, respectively, and $\theta_0$ and $\gamma$ are the projection angles of the impact parameter vector and velocity on the $\hat{z}$-axis, respectively.  The scheme's geometry is shown in Fig.~\ref{fig:external particle}, and a detailed derivation is presented in Appendix~\ref{app:detail_calc}.
    Figs.~\ref{fig:results-charged-int},~\ref{fig:results-neutral-int} show that the dephasing increases with increasing velocity and decreasing impact parameter.
    }
    \label{table:results}
\end{table*}

First, we discuss the dephasing for the charged micro-particle, discussed in Secs.~\ref{sec:charged}. They are plotted in Fig.~\ref{fig:results-charged-int}.
Here, we specify the experimental parameters and summarise the dipole moment orientation and evolution assumptions.
The dephasing that is plotted is given in Eqs.~\eqref{eq:gamma-cc},~\eqref{eq:gamma-cd(per)} and~\eqref{eq:gamma-cd(ind)} corresponding to dephasing in a charged interferometer. We will also discuss the possibility of the charged interferometer being an ideal quantum sensor to detect another electromagnetically charged particle moving in the vicinity.

The dephasing has been integrated in the frequency range from $\omega_\text{min}=2\pi/(t_f-t_i)$ to $\omega=\infty$. The experiment time is set to be $\tau =t_f-t_i = 4t_a + t_e$ (see figure~\ref{fig:Interferometer path}).
This frequency cut-off is chosen due to the frequency resolution of the discrete Fourier transform for the experimental setup. For the QGEM-specific example, we will take $\tau \sim 1$ s; for the CNOT gate, we will take $\tau\sim 1{\rm \mu s}$.

These parameters are chosen such that  no acceleration jitter can be averaged beyond the experimental time. We have also optimised  the projection angles, see Figs.~\ref{fig:charged-angles-01}-\ref{fig:charged-angles-10}, and the discussion in Appendix~\ref{app:proj_angles}. We have chosen the projection angles such that they  maximise the dephasing.  This will enable us to gauge the maximum acceleration noise the matter-wave interferometer can tolerate due to EM interactions. Also, it will help us do the total noise budget for the QGEM-type experiment~\cite{Bose:2017nin} and with the charged micro-particles~\cite{Barker:2022mdz}.

We have also plotted the dephasing given in Eqs.~\eqref{eq:gamma-dc(per)},~\eqref{eq:gamma-dc(ind)} 
and~\eqref{eq:gamma-dd} relating to the neutral micro-particle, which we will discuss below. While plotting this dephasing for the neutral micro-particle, we have made the following additional assumptions for the dipoles to obtain approximately the upper bounds on the dephasing. (1) In Eq.~\eqref{eq:gamma-cd(per)}, the external dipole is assumed to align approximately with the vector $\hat{\boldsymbol{r}}(t)$ during the time of the experiment $\tau$. 
(2) In Eq.~\eqref{eq:gamma-dc(per)} the interferometer dipole is assumed to align with the $\hat{z}$-axis.
(3) In Eq.~\eqref{eq:gamma-dd} we have assumed both the  above assumptions.

Other relevant parameters we have set are the maximum superposition size of $\Delta x = 20 \,\mu$m and $t_a = 0.50$ s and $t_e = 1.00$ s with total time $\tau=3$ s.
Furthermore, we have taken the time scale over which the noise is averaged, $T\sim b/v> 3$~s. We have also varied $T$ in Fig.\ref{fig:results-neutral-int}(a) for the neutral case for illustration. 

\subsection{Dephasing of a charged micro-particle}

As it is clear from the plots for the case of a charged micro-particle interferometer, the dephasing depends on the particle's velocity $v$, and minimum impact parameter, which we have taken to be $b=10-100\,\si{\micro\metre}$.
For our analysis to be valid, the ambient particle is expected to have a small velocity $ v\ll c$ since we have excluded relativistic effects in the potentials. 
In Fig. \ref{fig:results-charged-int}(a), the dephasing grows as the velocity increases for a fixed $b$. 
The charge-charge case dominates the dephasing over other cases. This can be seen from Fig. \ref{fig:results-charged-int}(d). The Coulomb interaction is by far the most dominant source of dephasing for the charged interferometer; see Figs.~\ref{fig:results-charged-int}(a-d). The charged interferometer can be treated as an excellent quantum sensor, sensing a charged ion moving with a velocity range of order $v\sim {\cal O}(10^{-1})\,{\rm m/s}$ in the vicinity of $b\sim 1$~m from the matter-wave interferometer (outside the range shown in the plot). 

The dephasing due to charge-dipole interaction is shown in Fig.~\ref{fig:results-charged-int}(b),(c).
Fig (b) shows the charged micro-particle interacting with a molecule with a dipole interaction, where the external dipole is intrinsic to the external particle, for example, in a water molecule.
The figure is plotted for 
a permanent dipole moment $\abs{\boldsymbol{d}_\text{ext}}\sim6.17\times10^{-30}\,\si{\coulomb\metre}$ of a water molecule~\cite{lide2004crc}, with $b=100\,\si{\micro\metre}$.
The dephasing is plotted as a function of the velocity of the external water molecule. For smaller velocities, the dephasing is small, but as the velocity increases, the dephasing increases and it also increases with the charge of the micro-particle in the interferometer. We can see that the dephasing is small compared to the dephasing due to the Coulomb interaction.

Fig.~\ref{fig:results-charged-int}(c) shows the dephasing due to the charged micro-particle and a dipole interaction where the external dipole is induced, for example, in  the polarized air molecules.
In this figure, we have taken the polarisability of di-nitrogen, ($1.710${\AA}$^3 = 1.903 \, \si{\ampere\squared\second^4\per\kilogram}$), which is most frequently present air molecule, and impact parameter $b=30\,\si{\micro\metre}$. Again, the dephasing is negligible.

We summarise our results in Fig.~\ref{fig:results-charged-int}(d), which shows the dephasings due to charge-charge (cc), charge-permanent dipole (cd(p)) and charge-induced dipole (cd(i)) as a function of the impact parameter $b$ for $v = 1 \,\si{\micro\metre\per\second}$.
The plot shows how the impact parameter $b$ influences dephasing in a fixed velocity scenario. 

\subsection{Dephasing of a neutral micro-particle}

\begin{figure*}[t!]
\centering
    \includegraphics[width=0.48\linewidth]{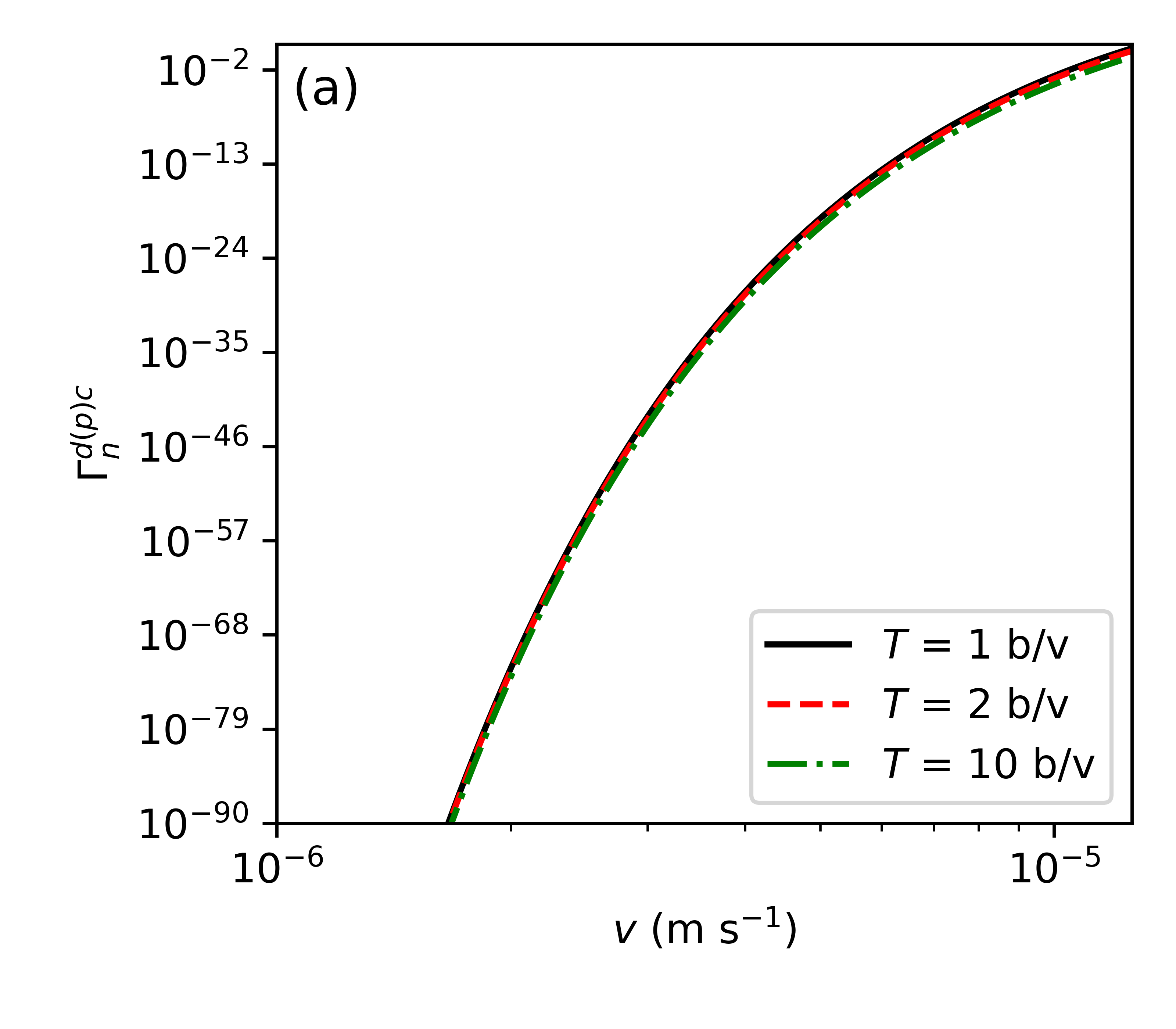}\hfill
    \includegraphics[width=0.48\linewidth]{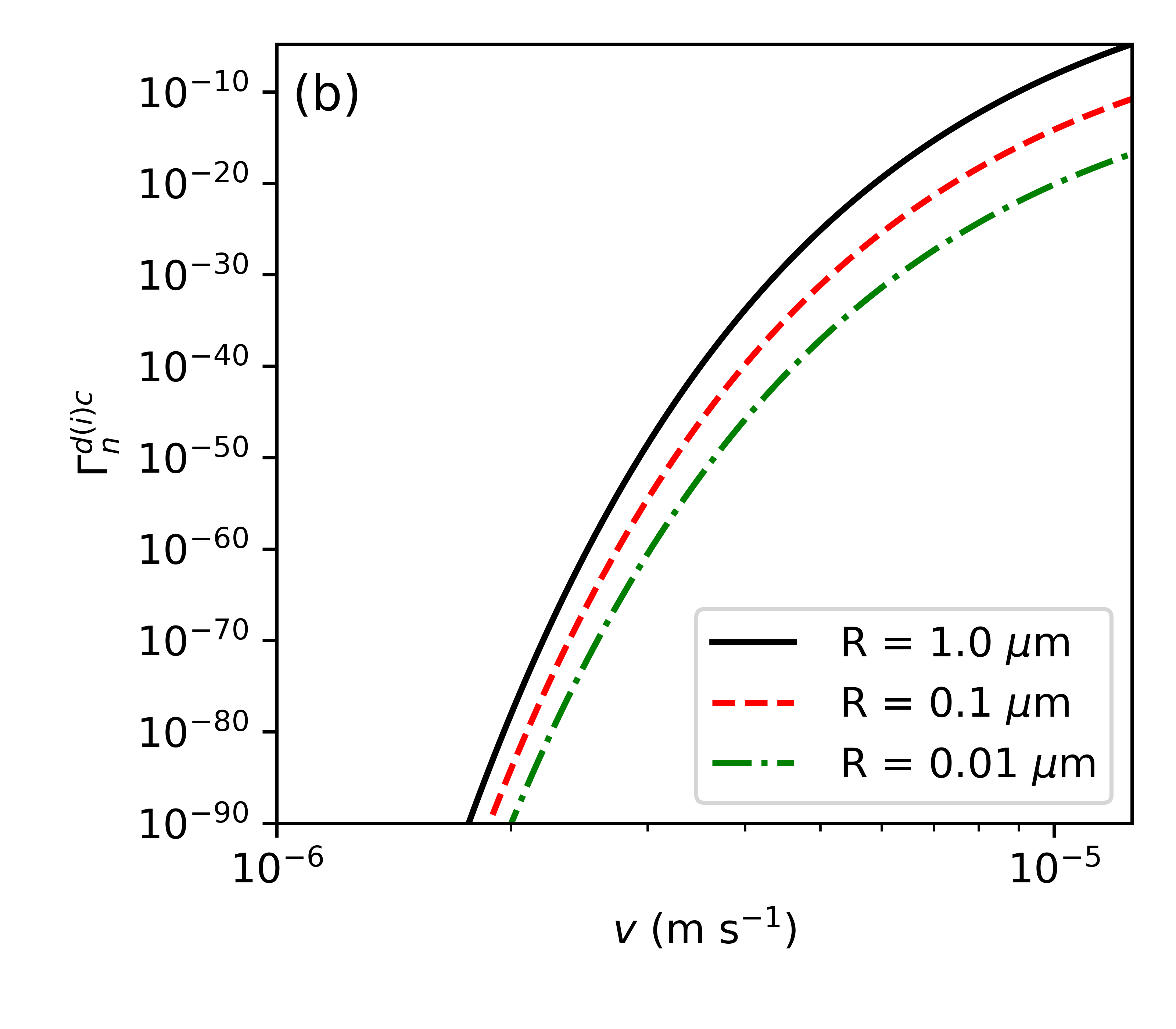}\par\medskip
    \includegraphics[width=0.48\linewidth]{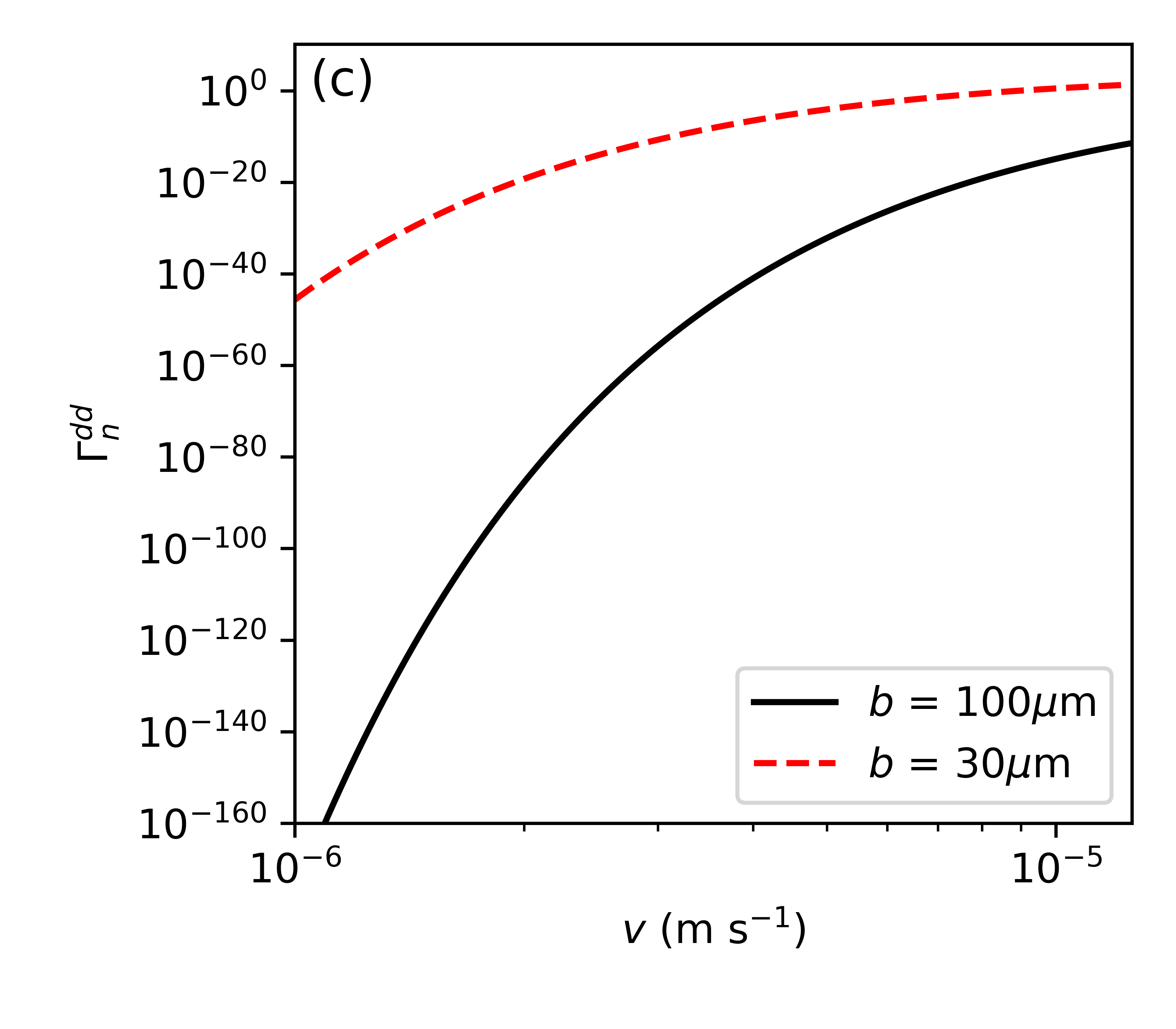}\hfill
    \includegraphics[width=0.48\linewidth]{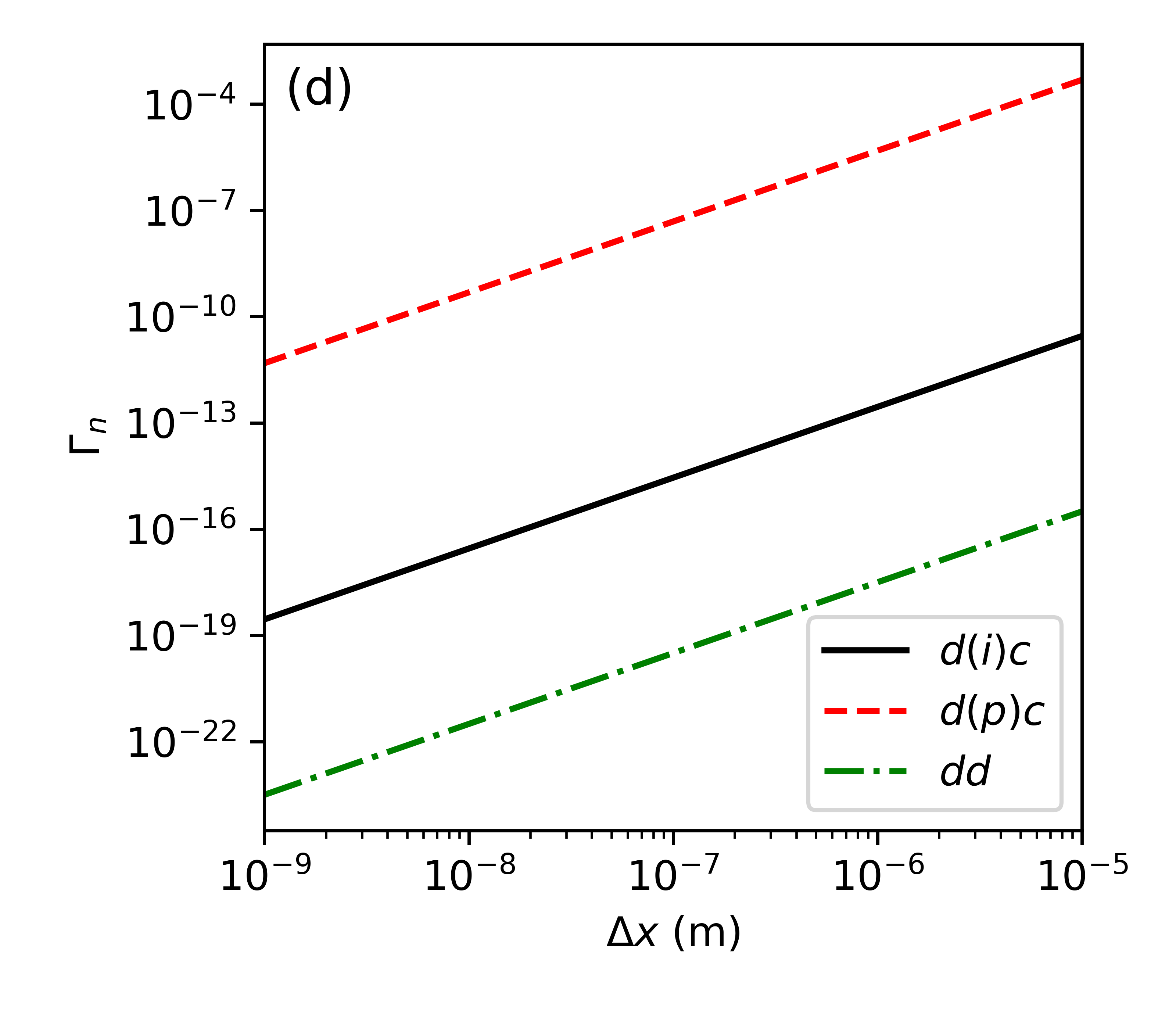}
\caption{Dephasing due to (a) the permanent dipole-charge interaction (d(p)c), (b) the induced dipole-charge interaction (d(i)c) and (c) the dipole-dipole interaction (dd) as a function of the velocity of the external particle.
The expressions summarised in Tab.~\ref{table:results} are plotted, where the integral is solved numerically.
With projection angles $\alpha=\beta=\gamma=\theta_0=\pi/4$ for Fig. (a), $\alpha=\beta=0$ for Fig.(b), and $\alpha=\theta_0 = \pi/4$, $\beta=0$ for Fig.(c), see Appendix~\ref{app:proj_angles}.
We take $\Delta x = 20{\rm \mu m}$ in all the cases.
In Fig.(a) $b=100\,\si{\micro\metre}$, $d_\text{int} = 0.1 e\,\si{\micro\metre}$ and $q_\text{ext} =e$. Different lines for the noise time scale $T$ are shown. There is not much noticeable difference in the dephasing rate due to different values of $T=b/v$. 
 In Fig.(b) $b=100\,\si{\micro\metre}$, $d_\text{int}$ is induced by the external charge $q_\text{ext} =e$ (with $\epsilon_r = 5.7$) and different lines for the interferometric particle's radius are plotted. 
In Fig.(c) $d_\text{int} = 0.1 e\,\si{\micro\metre}$ and the external particle is taken to be a water molecule, the dephasing is shown for different impact parameters. 
In Fig.(d), we summarise the dephasing from neutral micro-particles interactions as a function of the superposition width $\Delta x$, for $v = 10 \,\si{\micro\metre\per\s}$, $b=100\,\si{\micro\metre}$, and the interaction-specific parameters discussed above for the charge and dipoles. The permanent dipole of a matterwave interferometer interacting with an external charge gives the largest dephasing here. We have ensured that $T=b/v > \tau\sim 3$ s.}
\label{fig:results-neutral-int}
\end{figure*}

In the case of a neutral interferometer, the results are shown in Fig.~\ref{fig:results-neutral-int}.
Fig.~\ref{fig:results-neutral-int}(a) shows the dephasing due to the interaction of an intrinsic dipole of a micro-particle aligned at the $\hat{z}$-axis interacting with an external charge.
We have taken the external particle to have the charge of an electron.
Following the analysis in Appendix~\ref{app:proj_angles}, where we have chosen the angles to maximise the dephasing, the projection angles are considered $\alpha=\beta=\pi/2$.
Furthermore, as an example, we have taken the impact parameter $b=100\,\si{\micro\metre}$.
In this plot, we have also shown different values of the $T\sim b/v$ parameter to show how the dephasing behaves with varying interaction times. As the velocity increases for a fixed impact parameter, the dephasing increases. Nevertheless, the dephasing remains very small for our analysis to be valid $T=b/v> \tau=3$~s.

Fig.~\ref{fig:results-neutral-int}(b) shows the dephasing due to the interaction between an external charge and the dipole it induces within the diamond microsphere. We have varied the mass of the micro diamond by changing the radius from  $R\sim 1-0.01 {\rm \mu m}$, and the dielectric constant of the diamond is taken to be $\epsilon_r =5.7$.
The figure is very similar to (a), but the dephasing is significantly smaller (the plot is made for the impact parameter, $b=100\,\si{\micro\metre}$).
The dephasing would be further suppressed for smaller radii of the interferometer particle.
In this case, the maximal projection angles were $\alpha=\beta = 0$, which would maximise the dephasing; see Appendix~\ref{app:proj_angles}.

The dephasing due to the dipole-dipole interaction is given in fig.~\ref{fig:results-neutral-int}(c). 
Again, the interferometer is assumed to have a dipole aligned with the $\hat{z}$-axis of magnitude $\abs{\boldsymbol{d}_\text{int}}\sim 0.1 e\,\si{\micro\metre} \approx 1.6\times10^{-26}\,\si{\coulomb\metre}$.
The external dipole is assumed to be that of water, $\abs{\boldsymbol{d}_\text{ext}}\sim 6.17\times 10^{-30}\,\si{\coulomb\metre}$.
Following the analysis in Appendix~\ref{app:proj_angles} the projection angles are taken to be $\alpha = \beta=\pi/2$, which gives the maximum dephasing, e.g. worst-case scenario, if $vt/b<1$.
Different lines are shown for different impact parameters; one can see the dephasing is tiny.

It is worth pointing out that Ref.~\cite{Fragolino:2023agd} also studied the decoherence due to the dipole-dipole interactions in approximately the same parameter space. Specifically, the paper found the decoherence rate using the scattering model in the Born-Markov approximation. 
A direct comparison may not be valid, as the authors considered a quantum bath consisting of particles modelled as plane waves inside a box. In contrast, this paper considers a single external particle with classical trajectories.
Still, comparing our analysis with the analysis from~\cite{Fragolino:2023agd} is interesting. 
As the underlying assumptions are different, the predictions will, in general, differ, but a comparison nonetheless showed that when the impact factor $b$ from this analysis matches the size of the particle $R$ (and setting the number density corresponding to one environmental particle), the two predictions give the same order of magnitude prediction as expected (i.e., we match the length scale characterizing the distance between the system and environment to the same value in the two models).

Finally, Figs.~\ref{fig:results-neutral-int}(d) gives an overview of the dephasings described in Figs. (a)-(c) as a function of the superposition width, showing that a smaller superposition width reduces the dephasing.
The most dominant source of dephasing is from the permanent dipole interacting with an external (electron) charge for  $b=100\,\si{\micro\metre}$ and $v=10 \,\si{\micro\metre\per\second}$, which ensures $T\sim b/v > \tau$. The dephasing in all the cases is negligible.


\section{Applications}\label{sec:applications}

We have discussed above the dephasing expressions for the charged and the neutral interferometers due to external charges and dipoles. 
We now discuss the applications for both cases. 
The neutral interferometer is discussed in the context of the QGEM proposal for testing the quantum nature of gravity by witnessing the spin entanglement~\cite{Bose:2017nin}. 
We will also discuss the charged interferometer in the context of an ion-based quantum computer (where entanglement is due to the interaction of the photon between the two ions in adjacent ion traps) and study the dephasing due to the external charge and dipoles for a setup similar to the QGEM case.


\subsection{QGEM Protocol}\label{subsec:qgem}

\begin{figure}[b]
    \centering
    \includegraphics[width=\linewidth]{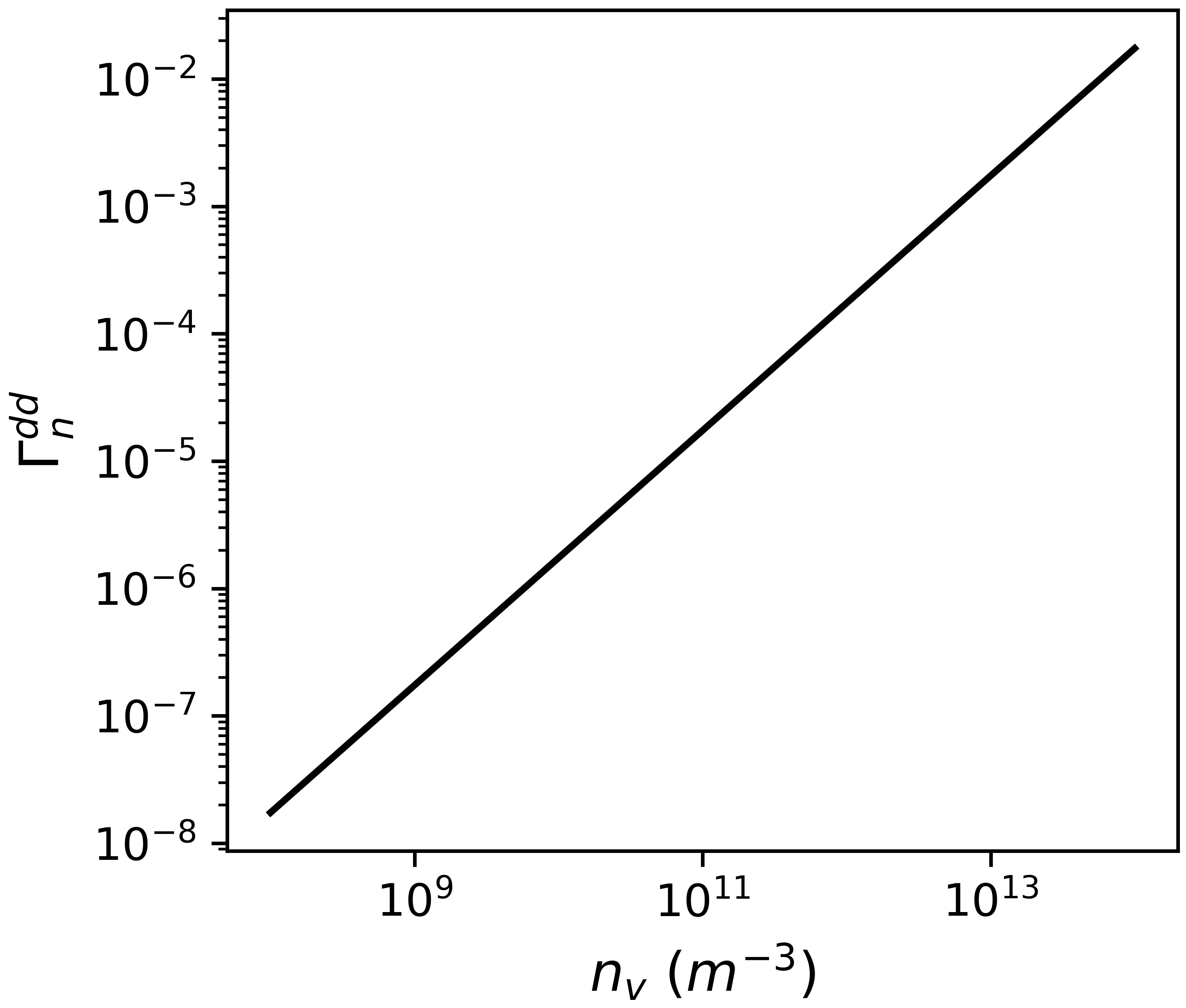}
    \caption{Dephasing for the QGEM microdiamond setup~\cite{Schut:2023hsy} as a function of number density $n_v\in[10^8\,\si{\per\metre\cubed},10^{14}\,\si{\per\metre\cubed}]\,$ in a square box of size $L=0.01\,\si{\metre}$.
    We consider the diamond to have a permanent dipole of $0.1\,e\,\si{\micro\metre}$, and mass $m= 10^{-15}$~kg. 
    The external gas particles are assumed to have a dipole similar to that of a water molecule with the distributions given in Eqs.~\eqref{eq:1}-\eqref{eq:4} and the temperature $T_\text{gas}=0.1\,\si{\milli\kelvin}$.
    We further assume that the maximum superposition size is $\Delta x=10~\mu\text{m}$ and that the total interferometric time is $\tau=1$~s.
    The dephasing is from the dipole-dipole interaction, see Fig.~\ref{fig:results-neutral-int}(c), now applied to the QGEM parameters and taking only the dominant frequency mode $\propto 1/\tau$ into account.}
    \label{fig:QGEM}
\end{figure}

In the context of the QGEM protocol, details on which can be found here~\cite{Bose:2017nin}, we consider two neutral interferometers that are set up in a way such their spatial superposition directions are parallel~\cite{Chevalier:2020uvv,Tilly:2021qef,Schut:2021svd,Schut:2023eux,Schut:2023hsy}.
For illustration, the paths of a single spatial interferometer are assumed to be as illustrated in figure~\ref{fig:Interferometer path}.
The initial spin state of the interferometers is given by:
\begin{equation}\label{WF}
    \ket{\psi(0)} = \frac{1}{2} \left( \ket{\uparrow}_1 + \ket{\downarrow}_1 \right)\otimes \left(\ket{\uparrow}_2 +  \ket{\downarrow}_2\right) \, ,
\end{equation}
which is a separable state (see Sec.~\ref{sec:RAN} for a short discussion on how the spin degrees of freedom are used to create spatial superposition states and recombine the interferometric paths).
A quantum gravitational interaction between the two massive interferometers will cause an entangling phase~\cite{Bose:2017nin,Marshman:2019sne,Bose:2022uxe,Danielson:2021egj,Carney_2019,Carney:2021vvt,Biswas:2022qto,Christodoulou:2022vte,Vinckers:2023grv,Elahi:2023ozf,Chakraborty:2023kel}. 

To simplify the analysis, we take the dephasing to be the same for both interferometers.
Taking into account the dephasing of the two interferometers, the final wavefunction at a time $t=\tau$ is described by:
\begin{align}
    \ket{\psi(\tau)} = \frac{e^{i\phi} e^{-i \delta 
    \phi}}{2} \big( & e^{i\delta 
    \phi} \ket{\uparrow}_1 \ket{\uparrow}_2 + e^{i\Delta\phi}\ket{\uparrow}_1 \ket{\downarrow}_2 \nonumber \\
    &+ e^{i\Delta\phi} \ket{\downarrow}_1 \ket{\uparrow}_2 + e^{- i \delta 
    \phi} \ket{\downarrow}_1 \ket{\downarrow}_2 \big) \label{eq:psi_final}
\end{align}
with
\begin{align}
    \phi &= \frac{\tau}{\hbar} \frac{G m^2}{d} \, , \qq{}
    \Delta\phi = \frac{\tau}{\hbar} \int_0^\tau \frac{G m^2}{\sqrt{d^2 + \Delta x^2(t)}} \,\dd{t} - \phi,
\end{align}
where $\delta \phi$ is the phase difference between the left and right arm of the interferometer due to interaction with the environmental particle, see Eq.~\eqref{eq:delta-phi-def}.

The density matrix, $\rho$, is given by the final wavefunction, $\rho = \ket{\psi}\bra{\psi}$. 
Averaging over the different runs of the experiment, we find $\mathbb{E}[\rho]$. 
Treating $\delta \phi$ as a statistical quantity with $\mathbb{E}[\delta\phi]=0$ as discussed in Sec.~\ref{sec:RAN}, and $\mathbb{E}[(\delta\phi)^2]=\Gamma_n$, see Eq.~\eqref{eq:GammaDef}, the averaged density matrix is given by~\footnote{Here we have used that $\mathbb{E}[e^{i\delta\phi}] \approx e^{-\mathbb{E}[(\delta\phi)^2]/2}$, assuming $\delta\phi$ follows a Gaussian distribution with $\mathbb{E}[\delta \phi] = 0$~\cite{van1992stochastic}.}:
\begin{equation}
\resizebox{0.5\textwidth}{!}{$
    \mathbb{E}[\rho] = \frac{1}{4}
    \begin{pmatrix}
        1 & e^{-\Gamma_n/2-i\Delta\phi} & e^{-\Gamma_n/2-i\Delta\phi} & e^{-2 \Gamma_n} \\
        e^{-\Gamma_n/2+i\Delta\phi} & 1 & 1 & e^{-\Gamma_n/2+i\Delta\phi} \\
        e^{-\Gamma_n/2+i\Delta\phi} & 1 & 1 & e^{-\Gamma_n/2+i\Delta\phi} \\
        e^{-2\Gamma_n} & e^{-\Gamma_n/2-i\Delta\phi} & e^{-\Gamma_n/2-i\Delta\phi} & 1
    \end{pmatrix}.$}
\end{equation}
$\Gamma_\text{n}$ is given by the results found in the previous sections for the neutral micro-particle. To witness the entanglement due to the gravitational interaction, we use the positive partial transpose (PPT) witness, which, in the case of two qubits, provides a sufficient and necessary condition for entanglement based on the Peres-Horodecki criterion~\cite{Horodecki:2009zz, Tilly:2021qef,Schut:2021svd}. 
The PPT witness is defined as $\mathcal{W} = \ket{\lambda_{-}}\bra{\lambda_{-}}^{T_2}$, where $\ket{\lambda_-}$ is the eigenvector corresponding to the lowest eigenvalue, $\lambda_-$, of the partially transposed density matrix and the superscript $T_2$ denotes the partial transpose over system $2$.
The witness for two interferometers aligned in a parallel way, corresponding to the wavefunction in Eq.~\eqref{eq:psi_final}, was found to be~\cite{Schut:2023hsy}:
\begin{equation}
    \mathcal{W} = \frac{1}{4} (1\otimes1 - \sigma_x\otimes\sigma_x + \sigma_z\otimes\sigma_y + \sigma_y\otimes\sigma_z ) \, ,
\end{equation}
with $\sigma_x, \sigma_y, \sigma_z$ the Pauli spin matrices.
The expectation value of the witness is given by~\cite{Chevalier:2020uvv,Tilly:2021qef,Schut:2021svd}:
\begin{equation}
    \langle \mathcal{W} \rangle = \Tr(\mathcal{W} \rho) = \lambda_{-}.
\end{equation}
For the wavefunction in Eq.~\eqref{eq:psi_final} the expectation value is found to be:
\begin{equation}
    \langle \mathcal{W} \rangle = \frac{1}{8}(1-e^{-2\Gamma_n}) - \frac{1}{2} \sin(\abs{\Delta\phi}) e^{-\Gamma_\text{n}/2} \, ,
\end{equation}
which reduces to the witness expectation value in Ref.~\cite{Schut:2023hsy} if $\Gamma_n=0$.
The entanglement detection condition is $\langle\mathcal{W}\rangle<0$. 
There is no entanglement detection if $\langle\mathcal{W}\rangle\geq0$.
Performing a short-time expansion of the witness, these inequalities can approximate the maximum allowed value of the dephasing rate:
\begin{equation}
    2 \Gamma_\text{n} - \frac{\abs{\Delta \phi}}{2} < 0.
\end{equation}
In paricular, if $\Gamma_\text{n} > \abs{\Delta\phi}/4$, then $\langle\mathcal{W}\rangle >0$ and there is no entanglement. Hence, the total dephasing due to the EM interaction must be $\Gamma_n < \abs{\Delta\phi}/4$ to detect the gravity-induced entanglement.

We now consider the induced dipole-charge and the dipole-dipole dephasing (Eqs.~\eqref{eq:gamma-dc(ind)} and~\eqref{eq:gamma-dd} respectively).
To simplify the analysis, we will consider the gas temperatures to be small ($T\sim0.1\si{\milli\kelvin}$) such that the Maxwell-Boltzmann distribution in Eq.~\eqref{eq:MBD} can be approximated with a Dirac-delta distribution around the velocity $\bar{v} = \sqrt{2 k_B T_\text{gas} /m_\text{gas}}$, with $m_\text{gas} = 4.8*10^{-26}\,\si{\kilogram}$ corresponding to the typical mass of an air molecule. 
We further simplify the analysis by considering only the dominant frequency mode $\omega=\omega_\text{min}$, allowing for an approximation of the Bessel function (see the last three rows of Table~\ref{table:results}).
These approximations are detailed and motivated in Appendix~\ref{app:approximations}. Furthermore, the timescale $T$ is taken to be larger than the experimental time, $T= 10\tau $, such that Eq.~\eqref{eq:Teq} holds.

In the context of the QGEM experiment, we consider the diamond to have a relative permittivity of $\epsilon_r = 5.1$~\cite{ibarra1997wide,Schut:2023hsy}, and mass $m= 10^{-15}$~kg. 
We further assume that the maximum superposition size is $\Delta x=10~\mu\text{m}$ and that the total time is $\tau=1$~s and the superposition is created following the scheme in Fig.~\ref{fig:Interferometer path}.
The transfer function of the interferometer is given Eq.~\eqref{eq:transferfunction}, where the times $t_e$, $t_a$ are scaled-down compared to previous figures such that the total experimental time is $1\,\si{\second}$.
For the external particle, we have taken the same properties as in fig.~\ref{fig:results-neutral-int}(a-d).

The dephasing from the dipole-dipole interaction for the QGEM experiment is shown in Fig.~\ref{fig:QGEM}, where we have assumed an external dipole of magnitude $\abs{\boldsymbol{d}_\text{ext}}=6.17*10^{-30}\,\si{\coulomb\metre}$, and the interferometer dipole is assumed to be $\abs{\boldsymbol{d}_\text{int}}= 10^{-7} e\,\si{\coulomb\metre}$.
The figure shows the total dephasing as a function of the number density of gas particles in a square box of size $L=0.01\,\si{\centi\metre}$ cooled to $T_\text{gas}=0.1\,\si{\milli\kelvin}$.
The dephasing in Fig.~\ref{fig:QGEM} considers multiple particles described by a distribution over velocities, impact parameters and projection angles. Although remnant charges would cause a deleterious dephasing in the QGEM experiment, remnant dipoles within the vacuum chamber for the number density shown in Fig.~\ref{fig:QGEM} have a dephasing such that the entanglement remains witnessable.  



\subsection{CNOT Gate}\label{subsec:cnot}

\begin{figure}[b]
    \centering
    \includegraphics[width=\linewidth]{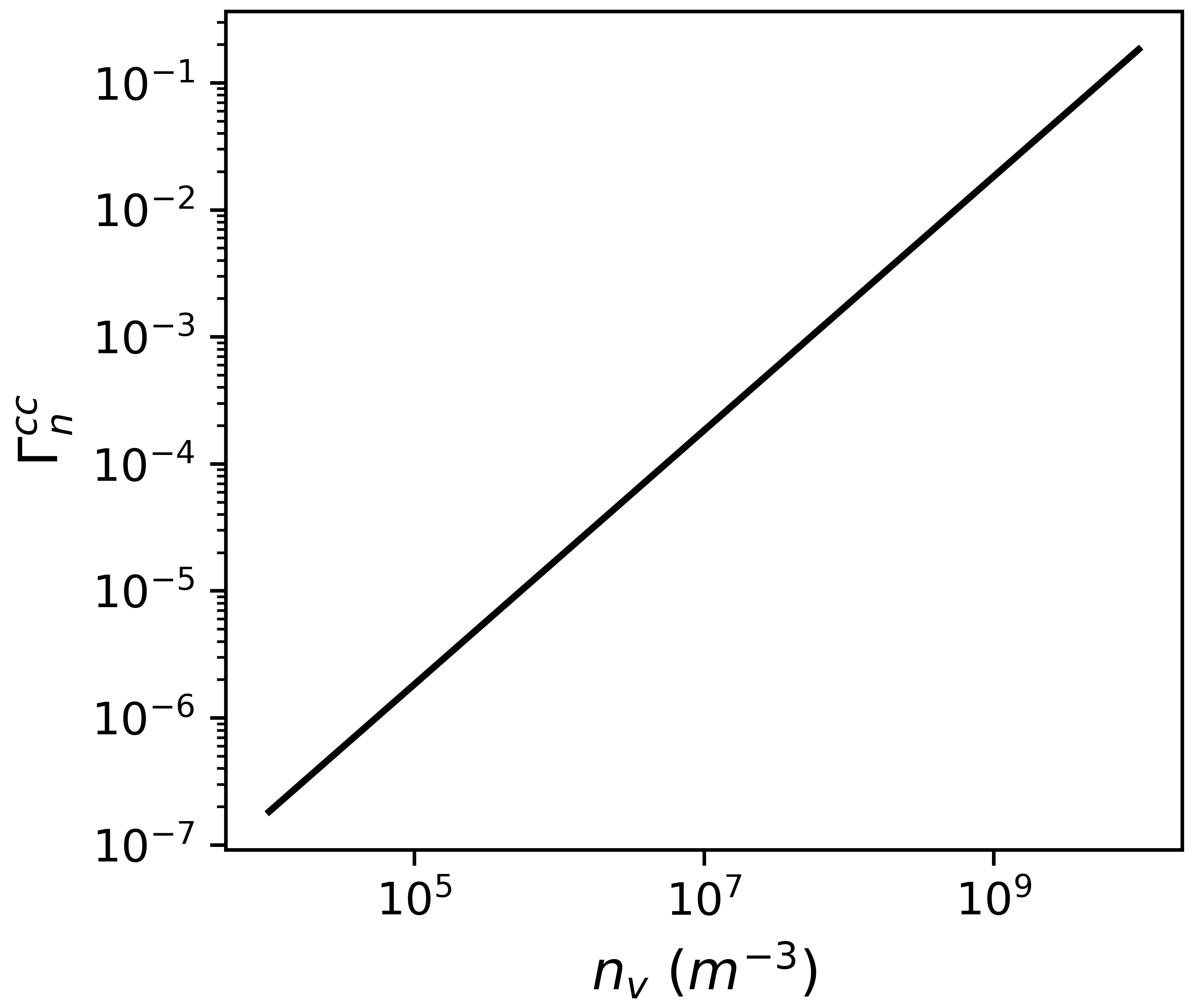}
    \caption{
    Charge-charge dephasing for ions of mass $m=10^{-27}\,\si{\kilogram}$ and total charge $q_\text{int}=e$ interacting with an external particle of charge $q_\text{ext}=10e$ as a function of the number density $n_v\in[10^4\,\si{\per\metre\cubed},10^{10}\,\si{\per\metre\cubed}]$ in a square box of size $L=0.01\,\si{\metre}$. The experimental time is $\tau = 1\,\si{\micro\second}$ and the superposition size is assumed to be $\Delta x=0.18 {\rm \mu m}$. The dephasing is approximated for temperatures $T_\text{gas}=0.1\,\si{\milli\kelvin}$ and $b_\text{min}\sim10^{-7}\,\si{\metre}$.}
    \label{fig:CNOT}
\end{figure}

In quantum computers, one often considers trapped charged particles~\cite{Wineland:1992,Wineland:1995}. Here, we consider a CNOT gate consisting of two trapped ions separated by a distance $d$. Their initial wavefunction will be assumed to be prepared in a product state consisting of individual spatial superpositions. The entanglement between the two trapped ions will then build up due to the EM interaction mediated via photon, essentially the Coulomb interaction. 
The final wavefunction of such an entangled system will be given by Eq.~\eqref{eq:psi_final}, with an entangled phase given by:
\begin{align}
    &\Delta\phi = - \frac{\tau}{\hbar} \int_0^\tau \frac{\kappa_e q_{1}q_{2}}{\sqrt{d^2 + \Delta x^2(t)}} \,\dd{t} - \phi\,, \\
    &\text{ with } \phi = - \frac{\tau}{\hbar} \frac{\kappa_e q_1q_2}{d} \, , 
\end{align}
where $\kappa_e = (4\pi \epsilon_0)^{-1}$, and $q_1$ and $q_2$ are internal charges of the ions in their respective traps.

In Fig.~\ref{fig:CNOT}, we show the dephasing $\Gamma^{cc}_{n}$ from the charge-charge Coulomb interaction. We have assumed that an external charge $q_\text{ext}=10\,e$ (which is not trapped) is flying by the ion traps. 
We assume for simplicity that each trapped ion has mass $m\sim 10^{-27}\,\si{\kilogram}$, charge $q_\text{int}=q_1=q_2=e$, and is trapped in a trap of frequency $\omega=10^{5}$ Hz. 
For concreteness, we can consider the size of the superposition size to be of the same order of magnitude as the zero point motion  $\Delta x = \sqrt{{\hbar}/{2m\omega}}$, which for our parameters gives $\Delta x\sim 0.18 {\rm \mu m}$. The entanglement builds up swiftly due to the Coulomb interaction,  which is many orders of magnitude larger than the gravitational interaction strength. The total experimental time we require is $\tau = 1 \,\si{\micro\second}$ for the entanglement phase to be order unity assuming a trap separation $d=50{\rm \mu m}$.
For illustration, we took a similar transfer function responsible for creating the spatial superposition in Fig.~\ref{fig:Interferometer path}.  

The dephasing in Fig.~\ref{fig:CNOT} has been found for multiple gas particles that have velocity, impact parameter and projection angles following the distributions in Eqs.~\eqref{eq:1}-~\eqref{eq:4}. 
The dephasing is approximated by imposing a minimal impact parameter $b\in[b_\text{min},L]$ with $b_\text{min}=10^3 R\sim10^{-7}\,\si{\metre}$ and $L=0.01\,\si{\metre}$, such that a large-argument expansion can be used to approximate the Bessel function.
This approximation is detailed in Appendix~\ref{app:approximations}.
Fig.~\ref{fig:CNOT} shows the dephasing as a function of the number density such that the dephasing is smaller than the entanglement phase.
Note that the simplifying assumptions for the CNOT gate are slightly different compared to the QGEM experiment because the two experimental setups have different experimental parameters, we refer to App.~\ref{app:approximations} for more detail.



\section{Conclusion} \label{sec:discussion}

In this paper, we have adapted techniques from investigating non-inertial and gravitational noise~\cite{Toros:2020dbf,Wu:2022rdv,Saulson:1984} to compute the dephasing arising from the electromagnetic interactions. 
We considered two distinct scenarios, one where the interferometer particle is charged and the other where the micron-size particle is neutral. The main formulae are summarized in Table~\ref{table:results}.

The charged interferometer can interact via Coulomb interaction with an external charged particle. 
The external particle's velocity is assumed to be constant, and its acceleration on both arms of the interferometer causes a jitter, which causes dephasing. 
The charged interferometer can also interact with a constantly moving dipole. 
The ambient particle can have a permanent dipole or an induced dipole. 
We considered both cases separately; see Figs.~\ref{fig:results-charged-int}(a-d). 
The largest dephasing occurs due to the Coulomb interaction, which sets the exclusion zone for an externally charged particle. 

In the neutral case, we have again four possibilities; a micro-particle is typically considered a diamond in our case, which has a dielectric property.  The micro diamond can then have a permanent and an induced dipole; hence, it can interact with an external charge and an external dipolar particle, see Figs.~\ref{fig:results-neutral-int}(a-d).
The external particle will impart a jitter in the trajectories due to the relative accelerations of the arms of the matterwave interferometer and, hence, will induce a dephasing.

Suppose two neutral micro-particles, each prepared in spatial superpositions and kept in adjacent locations as in the case of the QGEM experiment~\cite{Bose:2017nin}. In that case, the relative acceleration noise is present in both test masses. 
This will cause a dephasing, as shown in Fig.~\ref{fig:QGEM}, leading to entanglement degradation.
The approximate average dephasing was found considering the probability distributions in Eqs.~\eqref{eq:1}-\eqref{eq:4}.
We found that it requires $\Gamma_n < \abs{\Delta\phi}/4$, with $\Gamma_n$ denoting the dephasing and $2 \Delta\phi$ the entanglement phase, to witness entanglement in the QGEM setup (positive partial trace witness~\cite{Chevalier:2020uvv,Tilly:2021qef,Schut:2021svd}). In the QGEM-type experiment, the dephasing from the external jitters due to the dipole-dipole interactions is small enough for the parameter space considered in this paper. At the same time, the dipole-charge interactions cause too much jitter.

The relative acceleration noise can also affect the ion trap-based spatial qubits, forming a CNOT gate. We observed that our methodology could be successfully employed to characterize the electromagnetic noise for such a setup. 
We found that the average dephasing, considering the probability distributions in Eqs.~\eqref{eq:1}-\eqref{eq:4}, increases as shown in Fig.~\ref{fig:CNOT}. The developed methodology could thus be used to estimate the decoherence budget. 

However, our analysis has also its limitations, which is suggestive for a more refined modelling in future works. We were able to obtain analytical results by assuming that the trajectories of the external particles are approximately classical and that the dynamics is non-relativistic. We also supposed that the ensemble of external particles is characterized by flat distribution of angles, impact parameters, and a Gaussian distribution in the velocities. While the modelling seem to be plausible from the experimental perspective, the analysis could be further refined with the help of numerical packages as analytical calculations will likely become challenging. 

In summary, we have provided a way of characterizing electromagnetic sources of dephasing due to external particles in the context of matter-wave interferometry. The developed techniques can be applied in fundamental problems such as witnessing the quantum nature of gravity in a lab, and  it could find applications for the design of future ion-based quantum computers.


\section{Acknowledgements}
The Fundamentals of the Universe research program at the University of Groningen supports MS. MT acknowledges funding by the Leverhulme Trust (RPG-2020-197). S.B. would like to acknowledge EPSRC grants (EP/N031105/1, EP/S000267/1, and EP/X009467/1) and grant ST/W006227/1. 

\bibliography{references.bib}
\bibliographystyle{ieeetr}

\appendix

\section{Estimation of dephasing}\label{app:detail_calc}
This section shows the expressions for the Fourier transform of the acceleration, $a_x(\omega)$ and the corresponding PSD for the acceleration noise $S_{aa}(\omega)$, from which the dephasing due to the acceleration noise can be derived via 
\begin{equation}\label{eq:noise-def-app}
    \Gamma_{\rm n} = \frac{1}{2\pi}\frac{m^2}{\hbar^2} \int_{\omega_{\rm min}}^\infty S_{aa}(\omega)F(\omega)\dd{\omega} \, .
\end{equation}
with the transfer function $F(\omega)$ determined by the trajectory, see Eq.~\eqref{eq:transferfunction}. Since,
\begin{align}
    a(\omega) &= \int_{-\infty}^\infty a(t) e^{-i\omega t} \dd{t} \\
    a(t) &= \frac{1}{2\pi} \int_{-\infty}^\infty a(\omega) e^{i\omega t} \dd{\omega}
\end{align}
We rewrite the acceleration such that the Fourier transform can easily be performed:
\begin{equation}
    a(b\omega/v) = \frac{b}{v} \int_{-\infty}^\infty a(vt/b) e^{-i(b\omega/v) (vt/b)} \dd{(vt/b)} \, .
\end{equation}
The Fourier transform is of a function $f(vt/b)$ as defined below is given by:
\begin{equation}\label{eq:fourier2}
\begin{aligned}
    f(vt/b) &= \frac{(vt/b)^{2n}}{((vt/b)^2 + c^2)^{\nu+1/2}} \\
    f(b\omega/v) &= \frac{(-1)^n c^\nu \sqrt{\pi}}{2^\nu \Gamma(\nu+1/2)} \pdv[2n]{(b\omega/v)^\nu K_\nu(c\,b\omega/v)}{(b\omega/v)} 
\end{aligned}
\end{equation}
with $\Gamma(x)$ the Euler-Gamma-function~\footnote{
With relevant values $\Gamma(3/2) = \sqrt{\pi}/2$, $\Gamma(2)=1$, $\Gamma(5/2) = 3\sqrt{\pi}/4$ and $\Gamma(3)=2$.
} and $K_\nu$ the modified Bessel function of the second kind with order $\nu$.
The above Fourier transform holds only if $\Re(c)>0$ and $\Re(\nu+1/2)>n$.
The modified Bessel functions have the property that:
\begin{equation}
\begin{aligned}\label{eq:bessel-prop}
    \pdv{K_\nu(x)}{x} &= - \frac{1}{2} (K_{\nu-1}(x) + K_{\nu+1}(x)) \\
    K_{\nu-1}(x) &= K_{\nu+1}(x) - \frac{2\nu}{x} K_\nu(x)
\end{aligned}
\end{equation}
The modified Bessel functions with half-integers have the analytic formula:
\begin{equation}
\begin{aligned}
    K_{1/2}(u) &= \sqrt{\frac{\pi}{2u}} \mathrm{e}^{-u} \\
    K_{3/2}(u) &= \sqrt{\frac{\pi}{2u}} \left(1+\frac{1}{u}\right) \mathrm{e}^{-u} \\
    K_{5/2}(u) &= \sqrt{\frac{\pi}{2u}} \left(1+\frac{3}{u}+\frac{3}{u^2}\right) \mathrm{e}^{-u}
\end{aligned}
\end{equation}
The integer-modified Bessel functions of the second kind do not have an analytical formula but can be expressed in the integral form:
\begin{equation}
    \begin{aligned}
        K_0(u) &= \int_0^\infty \frac{\cos(t u) }{\sqrt{1+t^2}} \dd{t} \\
        K_1(u) &= \frac{1}{u} \int_0^\infty \frac{\cos(t u) }{(1+t^2)^{3/2}} \dd{t} \\
        K_2(u) &= \frac{3}{u^2} \int_0^\infty \frac{\cos(t u) }{(1+t^2)^{5/2}} \dd{t}
    \end{aligned}
\end{equation}

\subsection{Charge-charge interaction}
As presented in Eq.~\eqref{eq:ax-cc}, the acceleration due to the charge-charge interaction in the $\hat{x}$-direction is:
\begin{equation}
    a_x^\text{cc}(t) = \frac{q_\text{int}q_\text{ext}}{4\pi \epsilon_0 m b^2} \frac{\cos(\alpha) + (vt/b) \cos(\beta)}{(1+(vt/b)^2)^{3/2}}.
\end{equation}
Transforming the acceleration $a_x^\text{cc}(t)$ into the frequency domain, we use Eq.~\eqref{eq:fourier2} (with $c=1,\nu=1, n=0,1/2$) to obtain:
\begin{equation}\label{eq:FTmonopoleaccel}
\begin{aligned}
    a_x^\text{cc}(\omega) = \frac{q_\text{int} q_\text{ext} \omega}{4\pi \epsilon_0 m v^2}  \bigg[ &\cos(\alpha) K_{1} \left(\frac{b\omega}{v}\right) \\ &\,\,-i\cos(\beta)K_{0}\left(\frac{b\omega}{v}\right)\bigg]\,,
\end{aligned}
\end{equation}
where $K_{\nu}(x)$ is a modified Bessel function, and we used its properties in Eq.~\eqref{eq:bessel-prop}.
Note that after Fourier transform, the unit of $a_x(\omega)$ is $\rm m/(s^2\cdot Hz)$, which is because $a_x(t)=\int a_x(\omega)e^{i\omega t}d\omega$ has a unit $\rm m/s^2$ and $d\omega$ has a unit Hz.

Substituting Eq.~\ref{eq:FTmonopoleaccel} for the acceleration due to the Coulomb interaction into Eq.~\ref{eq:PSD_wienerkhinchin} for the PSD results in the desired PSD of the acceleration noise from the Coulomb interaction:
\begin{equation}\label{eq:Saacc-app}
\begin{aligned}
    S_{aa}^\text{cc}(\omega) = \frac{q_\text{int}^2 q_\text{ext}^2 \omega^2}{16\pi^2 T m^2 v^4 \epsilon_0^2} &\biggr[\cos^2(\alpha)K_{1}^2\left(\frac{b\omega}{v}\right)\\
    &+\cos^2(\beta)K_{0}^2\left(\frac{b\omega}{v}\right)\biggr].
\end{aligned}
\end{equation}
Note that the unit of $S_{aa}(\omega)=|a_x(\omega)|^2/T$ is $\rm m^2/(s^4\cdot Hz)$.
Using Eq.~\eqref{eq:noise-def-app} gives the dephasing in Eq.~\eqref{eq:gamma-cc}.

\subsection{Charge-dipole interaction}
\subsubsection{Permanent dipole}
As presented in Eq.~\eqref{eq:ax-cd(per)}, the acceleration due to the charge-permanent dipole interaction in the $\hat{x}$-direction is:
\begin{equation}
    a_x^\text{cd(p)}(t) = \frac{q\abs{\mathbf{d}_\text{ext}}}{2 \pi m b^3 \epsilon_{0}} \frac{\cos(\alpha)+(vt/b)\cos(\beta)}{(1+(vt/b)^2)^2}\,.
\end{equation}
Transforming the acceleration $a_x^\text{cd(p)}(t)$ into the frequency domain, we use Eq.~\eqref{eq:fourier2} (with $c=1,\nu=3/2, n=0,1/2$) to obtain:
\begin{equation}
\begin{aligned}
    a_x^\text{cd(p)}({\omega}) =  \frac{q|\mathbf{d}_\text{ext}|\omega^{3/2}}{4\pi\epsilon_0 mv^{3/2}b^{3/2}} \sqrt{\frac{\pi}{2}} &\frac{b}{v} \Bigg[\cos(\alpha) K_{3/2}\left(\frac{b\omega}{v}\right) \\
    & - i\cos(\beta) K_{1/2}\left(\frac{b\omega}{v}\right) \Bigg] \,.
\end{aligned}
\end{equation}
where $K_{\nu}(x)$ is a modified Bessel function, and we used its properties in Eq.~\eqref{eq:bessel-prop}.

Now substituting this expression into the definition of the PSD of the acceleration noise results in:
\begin{equation}
\begin{aligned}
     S_{aa}^\text{cd(p)}(\omega) = \frac{q^2|\mathbf{d}_\text{ext}|^2\omega^3}{32 \pi T m^2 v^5 b \epsilon_{0}^2} &\Biggr[\cos^2(\alpha) K_{3/2}^2\left(\frac{b\omega}{v}\right) \\ 
    & + \cos^2(\beta) K_{1/2}^2\left(\frac{b\omega}{v}\right) \Biggr].
\end{aligned}
\end{equation}
The resulting dephasing is given in Eq.~\eqref{eq:gamma-cd(per)}.

\subsubsection{Induced dipole}
As presented in Eq.~\eqref{eq:ax-cd(ind)}, the acceleration due to the charge-permanent dipole interaction in the $\hat{x}$-direction is:
\begin{equation*}
        a_x^\text{cd(i)}(t) =
        \frac{q^2_\text{int}\alpha_\text{pol}}{8 \pi^2 m b^5 \epsilon_{0}^2} \frac{\cos(\alpha)+(vt/b)\cos(\beta)}{(1+(vt/b)^2)^3} .
\end{equation*}
Performing a Fourier transform given in Eq.~\eqref{eq:fourier2} (with $c=1,\nu=5/2, n=0,1/2$) to find the acceleration in the frequency domain gives:
\begin{equation}
    \begin{aligned}
    a_x^\text{cd(i)}(\omega) = \frac{q^2\alpha_\text{pol} \omega^{5/2}}{64 \pi^2 m b^{3/2}v^{7/2} \epsilon_{0}^2} &\sqrt{\frac{\pi}{2}} \Biggr[\cos(\alpha)K_{5/2}\left(\frac{b\omega}{v}\right)\\
    & - i\cos(\beta)K_{3/2}\left(\frac{b\omega}{v}\right)\Biggr].
    \end{aligned}
\end{equation}
Substituting this expression into the definition of the PSD of the acceleration noise results in a slightly modified expression compared to the permanent dipole case:
\begin{equation}
    \begin{aligned}
    S_{aa}^\text{cd(i)}(\omega) = \frac{q^4\alpha^2_\text{pol} \omega^5}{8192 \pi^3 m^2 v^7 b^3 \epsilon_{0}^4 T} \Biggr[&\cos^2(\alpha)K^2_{5/2}\left(b\omega/v\right)\\
    & + \cos^2(\beta)K^2_{3/2}(b\omega/v)\Biggr].
    \end{aligned}
\end{equation}
The resulting dephasing is given in Eq.~\eqref{eq:gamma-cd(ind)}.

\subsection{Dipole-charge interaction}

\subsubsection{Permanent dipole}
As presented in Eq.~\eqref{eq:ax-dc(per)} the acceleration in the $\hat{x}$-direction due to the permanent dipole interacting with an external charge is:
\begin{align}
    a_x^\text{d(p)c}(t)
    = \frac{q_\text{ext}|\mathbf{d}_\text{int}|}{2 \pi m b^3 \epsilon_{0}} &\frac{\cos(\alpha)+(vt/b)\cos(\beta)}{(1+(vt/b)^2)^{5/2}} \\ \nonumber 
    &\bigg[ \cos(\theta_0) + (vt/b)\cos(\gamma)\bigg],
\end{align}
The second line comes from the resolution of the dot product as:
\begin{align}
    \boldsymbol{d}_\text{int}\cdot\hat{\boldsymbol{r}}(t) 
    &= 
    \abs{\boldsymbol{d}_\text{int}} \frac{b_z + v_z t}{r(t)} \, , \\
    &= \abs{\boldsymbol{d}_\text{int}} \frac{\cos(\theta_0) + (v t/b) \cos(\gamma)}{\sqrt{1+(vt/b)^2}} \, , \label{eq:d-c-angles}
\end{align}
where we have assumed that the direction of the dipole vector of the interferometer is constant, i.e. $\boldsymbol{d}_\text{int}$ is not time-dependent, and that it points in the $\hat{z}$-direction.
For example, because the microdiamond sphere is magnetically trapped~\cite{Marshman:2023nkh} and its rotational degrees of freedom are cooled to near its ground state~\cite{Deli__2020}, achieving good control of the system. 

In the second line above, Eq.~\eqref{eq:d-c-angles} we have written the dot product in terms of the angle $\theta_0$, which is the angle between the $\hat{z}$-axis and the vector $b$, and the angle $\gamma$ which gives the $\hat{z}$-component of the velocity:
\begin{equation}\label{eq:second-proj-angles}
    b_z = b \cos(\theta_0) \, , \qq{} v_z = v \cos(\gamma) \, .
\end{equation}
The Fourier transform of the acceleration $a_x^\text{d(p)c}(t)$ is found from Eq.~\eqref{eq:fourier2} (with $c=1,\nu=2, n=0,1/2,1$) and gives:
\begin{equation}
\resizebox{0.5\textwidth}{!}{$
\begin{aligned}
    a_x^\text{d(p)c}({\omega}) &=
    \Bigg[
    \cos(\alpha) 
    \bigg( K_{2}\left(\frac{b\omega}{v}\right) \cos(\theta_0) - i K_{1}\left(\frac{b\omega}{v}\right) \cos(\gamma) \bigg) \\
    &- i \cos(\beta) 
    \bigg( K_{1}\left(\frac{b\omega}{v}\right) \cos(\theta_0) - i \bigg[\frac{b\omega}{v} K_1\left(\frac{b\omega}{v}\right) \\ &- K_0\left(\frac{b\omega}{v}\right)\bigg] \cos(\gamma) \bigg)
    \Bigg] \frac{q_\text{ext}|\mathbf{d}_\text{int}|\omega^2}{6\pi\epsilon_0 m v^3} \,.
\end{aligned}
$
}
\end{equation}
Where we used Eqs.~\eqref{eq:fourier2} to find the Fourier transform and Eqs.~\eqref{eq:bessel-prop} to simplify the expression.

Now substituting this expression into the Eq.~\eqref{eq:saa_def} of the PSD of the acceleration noise results in:
\begin{equation}
\begin{aligned}
    S_{aa}^\text{d(p)c}(\omega) &= \frac{q_\text{ext}^2|\mathbf{d}_\text{int}|^2\omega^4}{36\pi^2\epsilon_0^2 m^2 v^6 T} 
    \Bigg[
    \cos^2(\alpha) 
    \bigg( K_{2}^2\left(\frac{b\omega}{v}\right) \\ &\cos^2(\theta_0) + K_{1}^2\left(\frac{b\omega}{v}\right) \cos^2(\gamma) \bigg) \\
    &+ \cos^2(\beta) 
    \bigg( K_{1}^2\left(\frac{\omega}{bv}\right) \cos^2(\theta_0) \nonumber \\
    & + \bigg[\frac{b\omega}{v} K_1\left(\frac{b\omega}{v}\right) - K_0\left(\frac{b\omega}{v}\right)\bigg]^2 \cos^2(\gamma) \bigg)
    \Bigg]  \,.
\end{aligned}
\end{equation}
The resulting dephasing from Eq.~\eqref{eq:noise-def-app} and~\eqref{eq:transferfunction} is given by:
\begin{equation}
\begin{aligned}
    \Gamma_{\operatorname{n}}^\text{d(p)c} &= 
    \frac{4 q_\text{ext}^2|\mathbf{d}_\text{int}|^2 (\Delta x)^2}{9\pi^3\epsilon_0^2 \hbar^2 v^6 t_a^4}
    \int^\infty_{\omega_\text{min}} \frac{1}{\omega^2} \sin^4\left(\frac{t_a \omega}{2}\right) \\
    & \Bigg[
    \cos^2(\alpha) 
    \bigg( K_{2}^2\left(\frac{b\omega}{v}\right) \cos^2(\theta_0) \\ &\qq{} + K_{1}^2\left(\frac{b\omega}{v}\right) \cos^2(\gamma) \bigg) \\
    &+ \cos^2(\beta) 
    \bigg( K_{1}^2\left(\frac{b\omega}{v}\right) \cos^2(\theta_0) + \bigg[\frac{b\omega}{v} K_1\left(\frac{b\omega}{v}\right) \\ &- K_0\left(\frac{b\omega}{v}\right)\bigg]^2 \cos^2(\gamma) \bigg)
    \Bigg] \sin^2\left(\frac{\omega}{2}(2t_a+t_e)\right)
    \dd{\omega}
\end{aligned}
\end{equation}
Which has been presented in Eq.~\eqref{eq:gamma-dc(per)} in a simplified form. 

\subsubsection{Induced dipole}
Eq.~\eqref{eq:ax-dc(ind)} gave the acceleration in the $\hat{x}$-direction for the interaction between an external charge and the dipole it induces in the interferometer:
\begin{equation}
\begin{aligned}
        a_x^\text{d(i)c}(t)
        &= \frac{\epsilon_r-1}{\epsilon_r+2} \frac{q_\text{ext}^2 R^3}{2 \pi \epsilon_{0} m b^5 } \frac{\cos(\alpha)+(vt/b)\cos(\beta)}{(1+(vt/b)^2)^3}
\end{aligned}
\end{equation}
Performing a Fourier transform to find the acceleration in the frequency domain gives (from Eq.~\eqref{eq:fourier2} with $c=1,\nu=5/2, n=0,1/2$):
\begin{align}
    a_x^\text{d(i)c}(\omega) = &\frac{\epsilon_r-1}{\epsilon_r+2} \frac{q_\text{ext}^2 R^3 \omega^{5/2}}{16 \pi m b^{5/2}v^{5/2} \epsilon_{0}} \frac{b}{v}\sqrt{\frac{\pi}{2}} \nonumber \\
    &\Biggr[\cos(\alpha)K_{5/2}\left(\frac{b\omega}{v}\right)  - i\cos(\beta)K_{3/2}\left(\frac{b\omega}{v}\right)\Biggr].
\end{align}
Substituting this expression into the PSD of the acceleration noise results in a slightly modified expression compared to the permanent dipole case:
\begin{equation}\label{eq:Saadc-app}
    \begin{aligned}
    S_{aa}^\text{d(i)c} = &\left(\frac{\epsilon_r-1}{\epsilon_r+2}\right)^2 \frac{q_\text{ext}^4 R^6 \omega^5}{512 \pi \epsilon_{0}^2 m^2 v^7 b^3  T} \\ &\Biggr[\cos^2(\alpha)K^2_{5/2}\left(\frac{b\omega}{v}\right) + \cos^2(\beta)K^2_{3/2}\left(\frac{b\omega}{v}\right)\Biggr]\,,
    \end{aligned}
\end{equation}
from which Eq.~\eqref{eq:gamma-dc(ind)} is derived.

\subsection{Dipole-dipole interaction}
The acceleration due to dipole-dipole interactions is given by~\cite{griffiths2013introduction}:
\begin{equation}
    \boldsymbol{a}_\text{dd} 
    = \frac{3 \,\hat{\boldsymbol{r}}}{4\pi\epsilon_0 m} \left[ \frac{\boldsymbol{d}_\text{ext} \cdot \boldsymbol{d}_\text{int}}{r^4(t)} - 3 \frac{ (\boldsymbol{d}_\text{ext} \cdot \hat{\boldsymbol{r}}) (\boldsymbol{d}_\text{int}\cdot \hat{\boldsymbol{r}})}{r^4(t)} \right] 
\end{equation}
To resolve the dot products, we recall from previous sections the assumption we have made that 
\begin{enumerate}
    \item The external particle has low velocity, meaning that $\boldsymbol{d}_\text{ext} \cdot \hat{\boldsymbol{r}}$ is approximately constant in time. 
    Furthermore, in the worst-case scenario $\boldsymbol{d}_\text{ext} \cdot \hat{\boldsymbol{r}} = \abs{\boldsymbol{d}_\text{ext}}$.
    \item The dipole of the interferometer particle is aligned with the $\hat{z}$-axis. Such that $\boldsymbol{d}_\text{int} \cdot \hat{\boldsymbol{r}}$ is as in Eq.~\eqref{eq:d-c-angles}.
    Assuming that the external particle has small velocity, $\boldsymbol{d}_\text{int} \cdot \hat{\boldsymbol{r}} \approx \abs{\boldsymbol{d}_\text{int}} \cos(\theta_0)$.
\end{enumerate}
From the orientation of the dipoles in these assumptions, it follows that:
\begin{enumerate}[resume]
    \item The angle between the interferometer and external dipoles is $\theta_0$, such that $\boldsymbol{d}_\text{int} \cdot \boldsymbol{d}_\text{ext} = \abs{\boldsymbol{d}_\text{int}} \abs{\boldsymbol{d}_\text{ext}} \cos(\theta_0)$.
\end{enumerate}
In this scenario, the acceleration in the $\hat{x}$-direction becomes:
\begin{align*}
    \boldsymbol{a}^\text{dd}_x(t) 
    &= - \frac{6}{4\pi\epsilon_0 m} \frac{\abs{\boldsymbol{d}_\text{ext}} \abs{\boldsymbol{d}_\text{int}} \cos(\theta_0)}{r^4(t)} \frac{\boldsymbol{r}_x}{r} \\
    &=  - \frac{6 \abs{\boldsymbol{d}_\text{ext}} \abs{\boldsymbol{d}_\text{int}} \cos(\theta_0)}{4\pi\epsilon_0 m b^4} \frac{\cos(\alpha) + (vt/b) \cos(\beta)}{(1+(vt/b)^2)^{5/2}}
\end{align*}
as presented in Eq.~\eqref{eq:ax-dd}.
Using the Fourier transform in Eq.~\eqref{eq:fourier2} and the properties given in Eq.~\eqref{eq:bessel-prop} gives:
\begin{align}
    a_x^\text{dd}(\omega) = &\frac{2 \abs{\boldsymbol{d}_\text{ext}} \abs{\boldsymbol{d}_\text{int}} \cos(\theta_0)}{4\pi\epsilon_0 m} \frac{\omega^2}{b v^3} \nonumber \\ &\Bigg[ \cos(\alpha) K_2\left(\frac{b \omega}{v}\right) - i \cos(\beta) K_1\left(\frac{b \omega}{v}\right) \Bigg]
\end{align}
which gives the PSD of the acceleration noise:
\begin{align}
    S_{aa}^\text{dd}(\omega) = &\frac{\abs{\boldsymbol{d}_\text{ext}}^2 \abs{\boldsymbol{d}_\text{int}}^2 \cos^2(\theta_0)}{4\pi^2\epsilon_0^2 m^2 T} \frac{\omega^4}{b^2 v^6} \nonumber \\ &\Bigg[ \cos^2(\alpha) K_2^2\left(\frac{b \omega}{v}\right) + \cos^2(\beta) K_1^2\left(\frac{b \omega}{v}\right) \Bigg] \label{eq:Saadd-app}
\end{align}
Together with the transfer function, this gives the dephasing due to the dipole-dipole interaction, as given in Eq.~\eqref{eq:gamma-dd}.

\section{Projection angles analysis}\label{app:proj_angles}
\begin{figure*}
\begin{minipage}[b]{0.31\textwidth}
  \includegraphics[width=\linewidth]{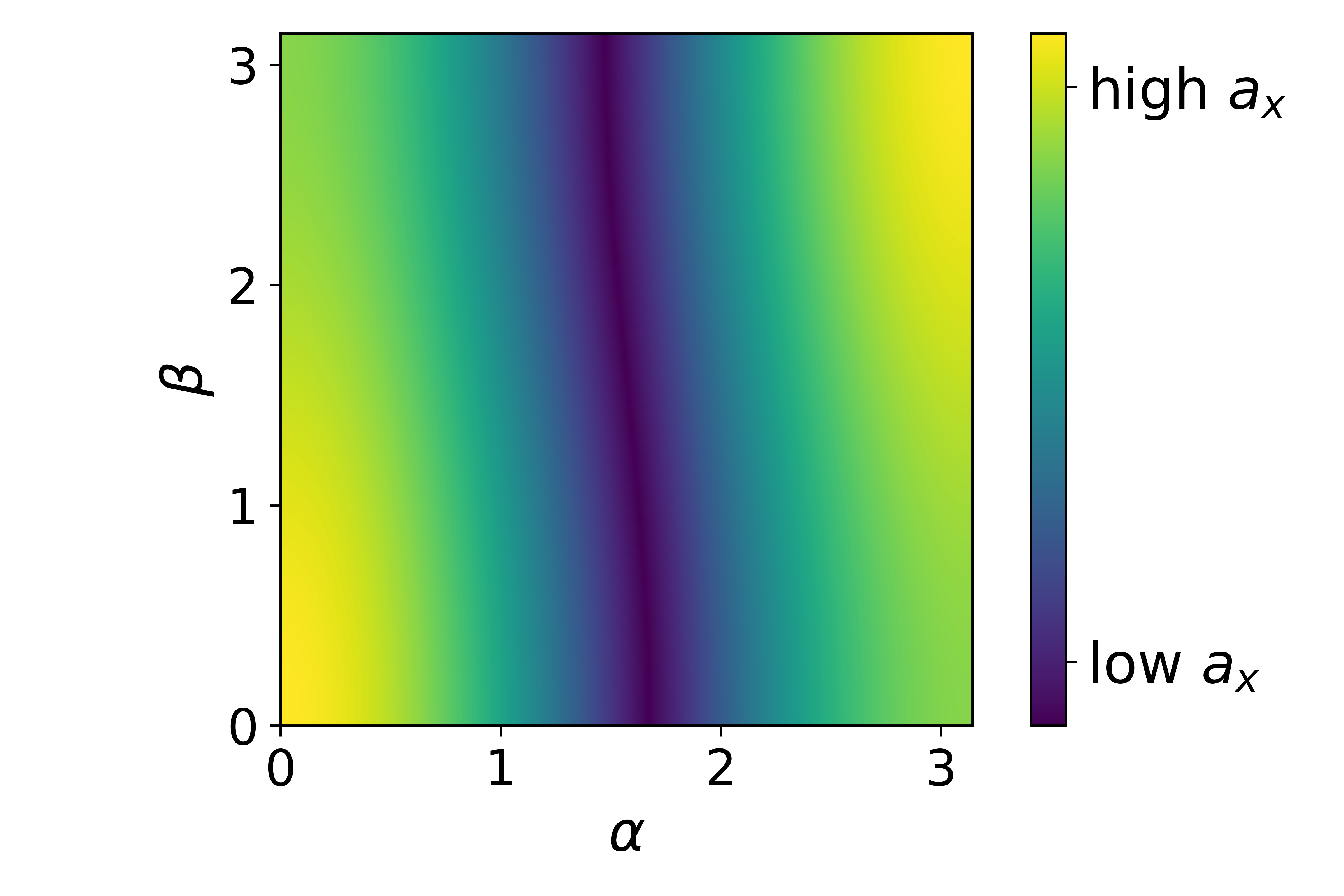}
  \caption{For $vt/b = 0.1$ the absolute value of the acceleration (Eq.~\eqref{eq:acc-angles} with $a=3$) as a function of the projection angles $\alpha, \beta \in [0,\pi]$.}
  \label{fig:charged-angles-01}
\end{minipage}%
\hfill 
\begin{minipage}[b]{0.31\textwidth}
  \includegraphics[width=\linewidth]{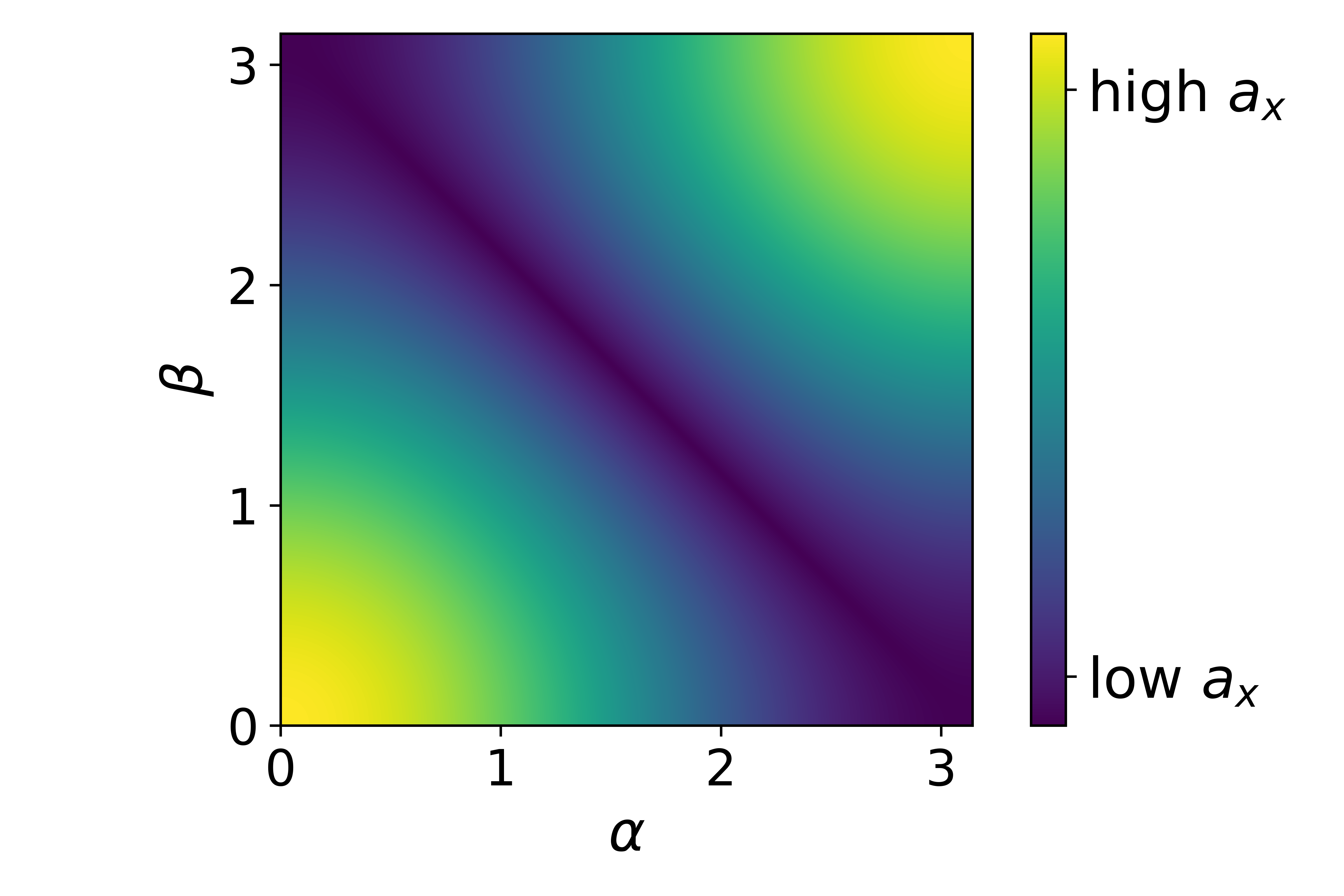}
  \caption{For $vt/b = 1$ the absolute value of the acceleration (Eq.~\eqref{eq:acc-angles} with $a=3$) as a function of the projection angles $\alpha, \beta \in [0,\pi]$.}
  \label{fig:charged-angles-1}
\end{minipage}%
\hfill
\begin{minipage}[b]{0.31\textwidth}
  \includegraphics[width=\linewidth]{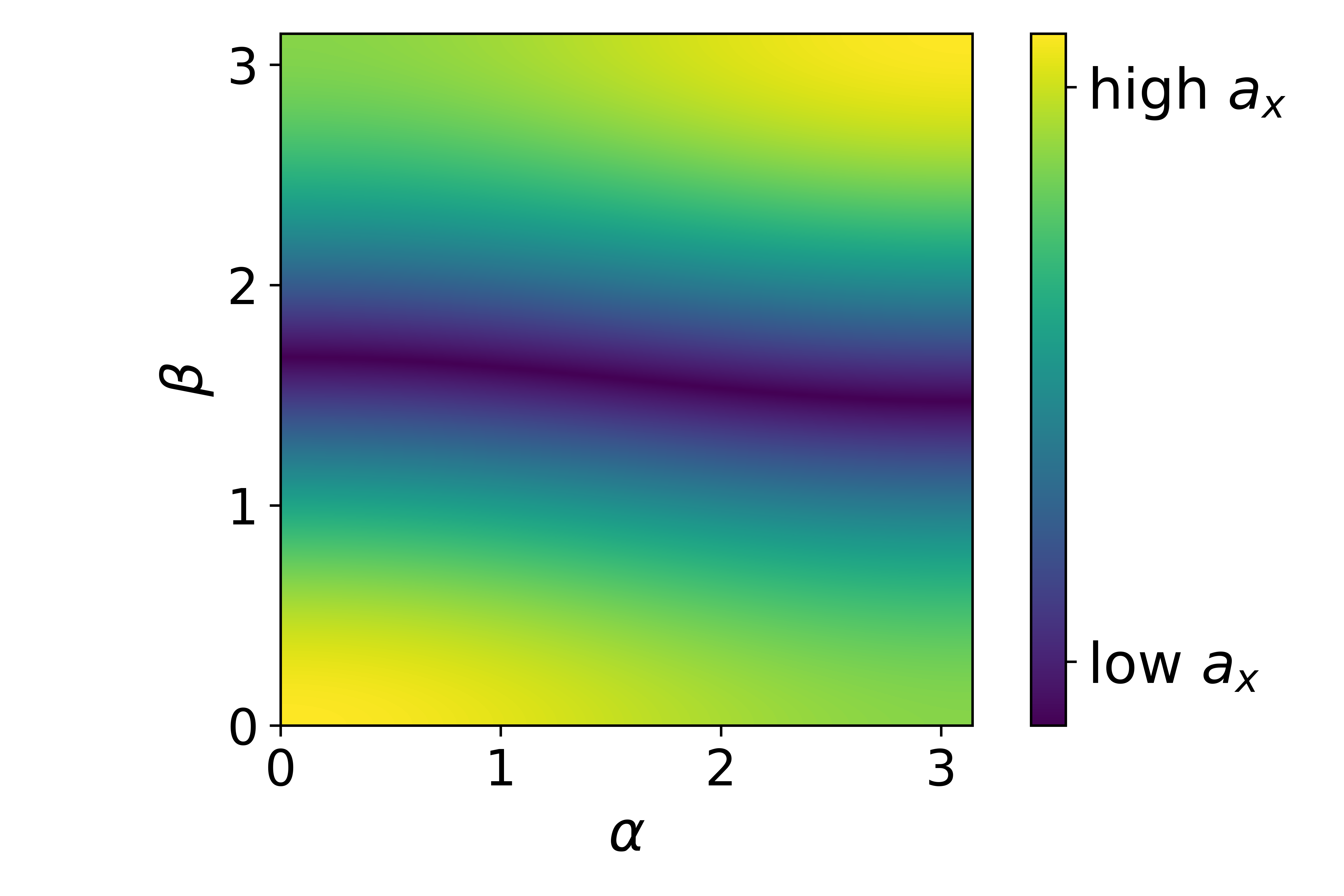}
  \caption{For $vt/b = 10$ the absolute value of the acceleration (Eq.~\eqref{eq:acc-angles} with $a=3$) as a function of the projection angles $\alpha, \beta \in [0,\pi]$.}
  \label{fig:charged-angles-10}
\end{minipage}%
\end{figure*}
To maximize the dephasing, we take the initial conditions as given by the projections angels $\alpha$ and $\beta$ (respectively, b and $v$ in the $\hat{x}$-direction) and in the case of an interferometer dipole $\theta_0$ and $\gamma$ (respectively, b and $v$ in the $\hat{z}$-direction), such that the absolute value of the acceleration in the $\hat{x}$-direction is maximal.

By maximising the acceleration, the noise is also maximised since the transfer function $S_{aa}(t) \propto a(t)^2$ and $\Gamma_\text{n} \propto S_{aa}(\omega) F(\omega)$, where the transfer function $F(\omega)$ is independent of the projection angles.
Physically, it is clear that a larger acceleration causes a larger dephasing.

\subsection{Charged interferometer}

For a charged interferometer, the relevant angles that determine the movement of the external particle are $\alpha$ and $\beta$; see Eq.~\eqref{eq:projectionangles}. These angles are maximised to find an upper bound on the dephasing. The figs.~\ref{fig:charged-angles-01}-\ref{fig:charged-angles-10} show the acceleration
\begin{equation}\label{eq:acc-angles}
    a(t) = \abs{C \frac{\cos(\alpha) + (vt/b) \cos(\beta)}{(1+(vt/b)^2)^a}}
\end{equation}
as a function of the angles $\alpha$ and $\beta$.
The magnitude of the acceleration is not given since it depends on the constant $C$, which is determined by the type of interaction (charged-charged, charged-permanent dipole or charged-induced dipole).
The value $a$ is also specified by the type of interaction, $a=3/2,2,3$, see Eqs.~\eqref{eq:ax-cc},~\eqref{eq:ax-cd(per)},~\eqref{eq:ax-cd(ind)}.

The figures show that for $vt > b$ by at least one order of magnitude, the acceleration is maximised for $\beta = 0,\pi$ and $\alpha$ arbitrarily.
If $v t < b$ by at least one order of magnitude, then the plot rotates ninety degrees, and the maximum value of the acceleration is given by $\alpha = 0,\pi$ and $\beta$ arbitrary. 
For $vt \sim b$ the maximisation is given by $\alpha =\beta = 0$ or $\alpha=\beta=\pi$.
The Figs.~\ref{fig:charged-angles-01}-\ref{fig:charged-angles-10} are approximately constant for $a=3/2,2,3$.

Therefore a simple way to choose the angles such that the absolute acceleration from the interaction is maximised is to choose $\alpha = \beta = 0$ (or $\pi$)~\footnote{Note that for the external dipole in section~\ref{sec:charged}, we have assumed that the particle has low velocity such that $\abs{\boldsymbol{r}\cdot\boldsymbol{d}}(t) \approx \abs{\boldsymbol{r}\cdot\boldsymbol{d}}$. 
So we have already made the assumptions on the magnitude of the velocity.}.

\subsection{Neutral interferometer}

\begin{figure*}
\hfill
\begin{minipage}[b]{0.40\textwidth}
  \includegraphics[width=\linewidth]{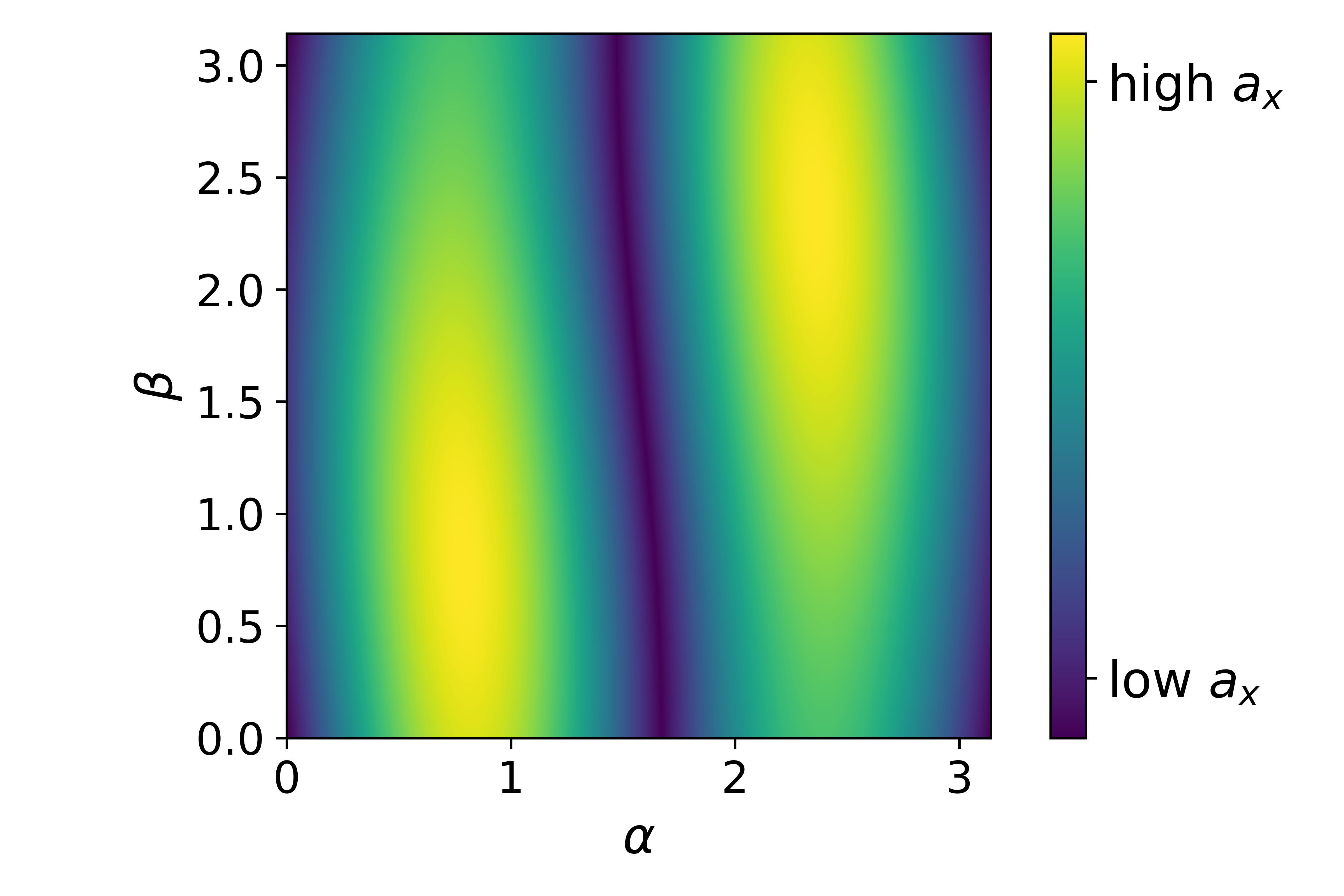}
  \caption{For $vt/b = 0.1$ the absolute value of the acceleration (Eq.~\eqref{eq:acc-angles-neutral}) as a function of the projection angles $\alpha, \beta \in [0,\pi]$.}
  \label{fig:neutral-angles-01}
\end{minipage}%
\hfill 
\begin{minipage}[b]{0.40\textwidth}
  \includegraphics[width=\linewidth]{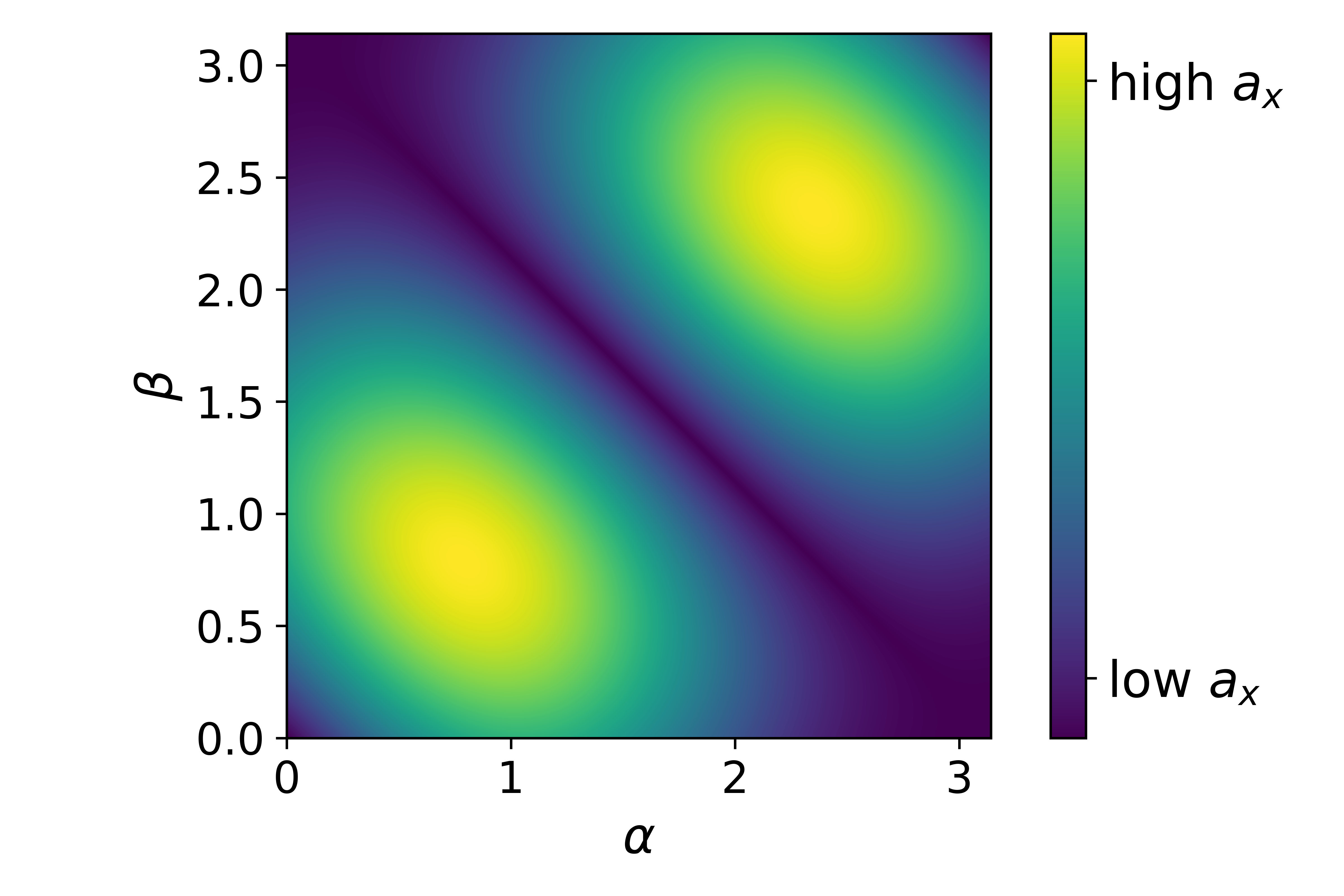}
  \caption{For $vt/b = 1$ the absolute value of the acceleration (Eq.~\eqref{eq:acc-angles-neutral}) as a function of the projection angles $\alpha, \beta \in [0,\pi]$.}
  \label{fig:neutral-angles-1}
\end{minipage}%
\hfill
\\
\hfill
\begin{minipage}[b]{0.40\textwidth}
  \includegraphics[width=\linewidth]{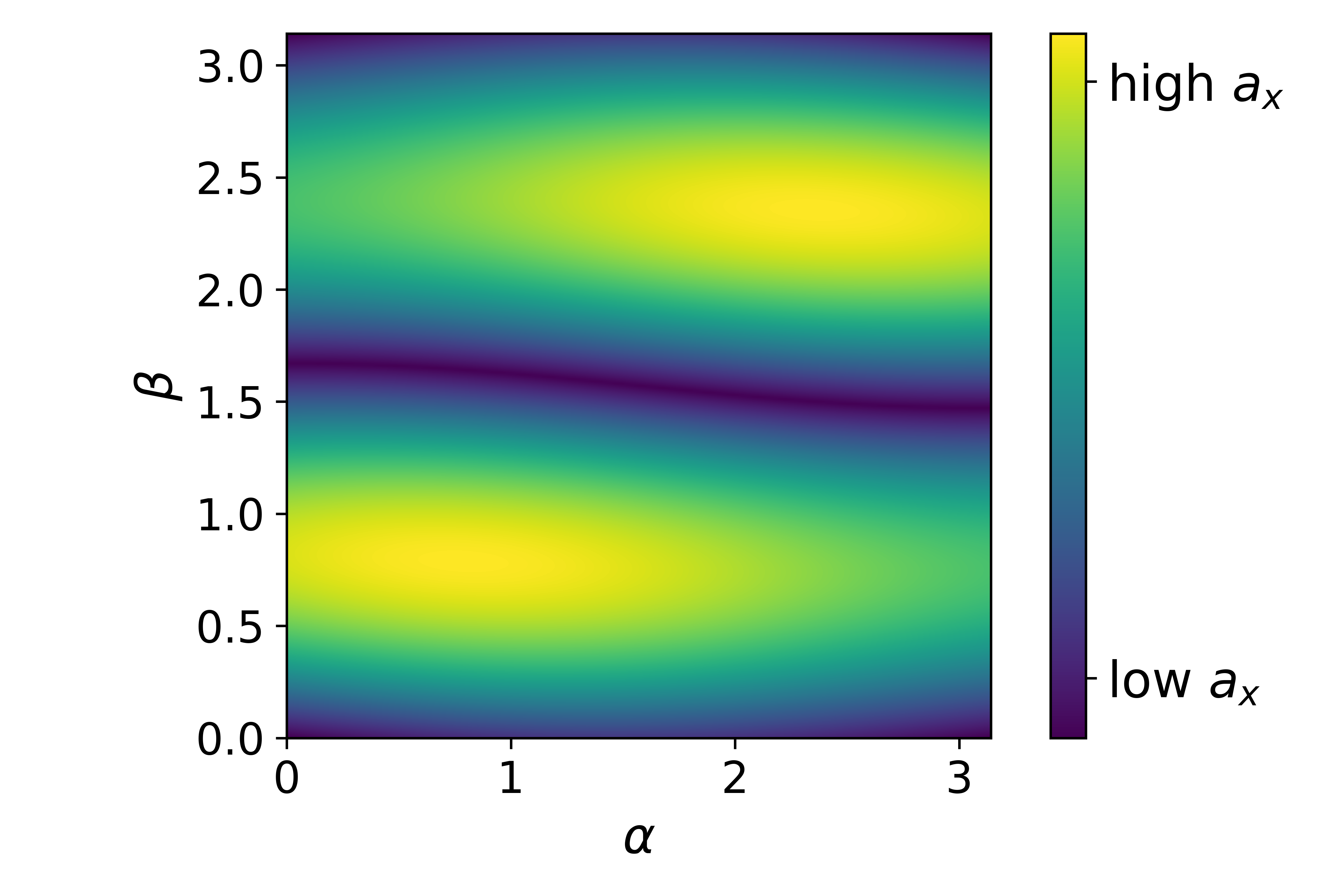}
  \caption{For $vt/b = 10$ the absolute value of the acceleration (Eq.~\eqref{eq:acc-angles-neutral}) as a function of the projection angles $\alpha, \beta \in [0,\pi]$.}
  \label{fig:neutral-angles-10}
\end{minipage}%
\hfill
\begin{minipage}[b]{0.40\textwidth}
  \includegraphics[width=\linewidth]{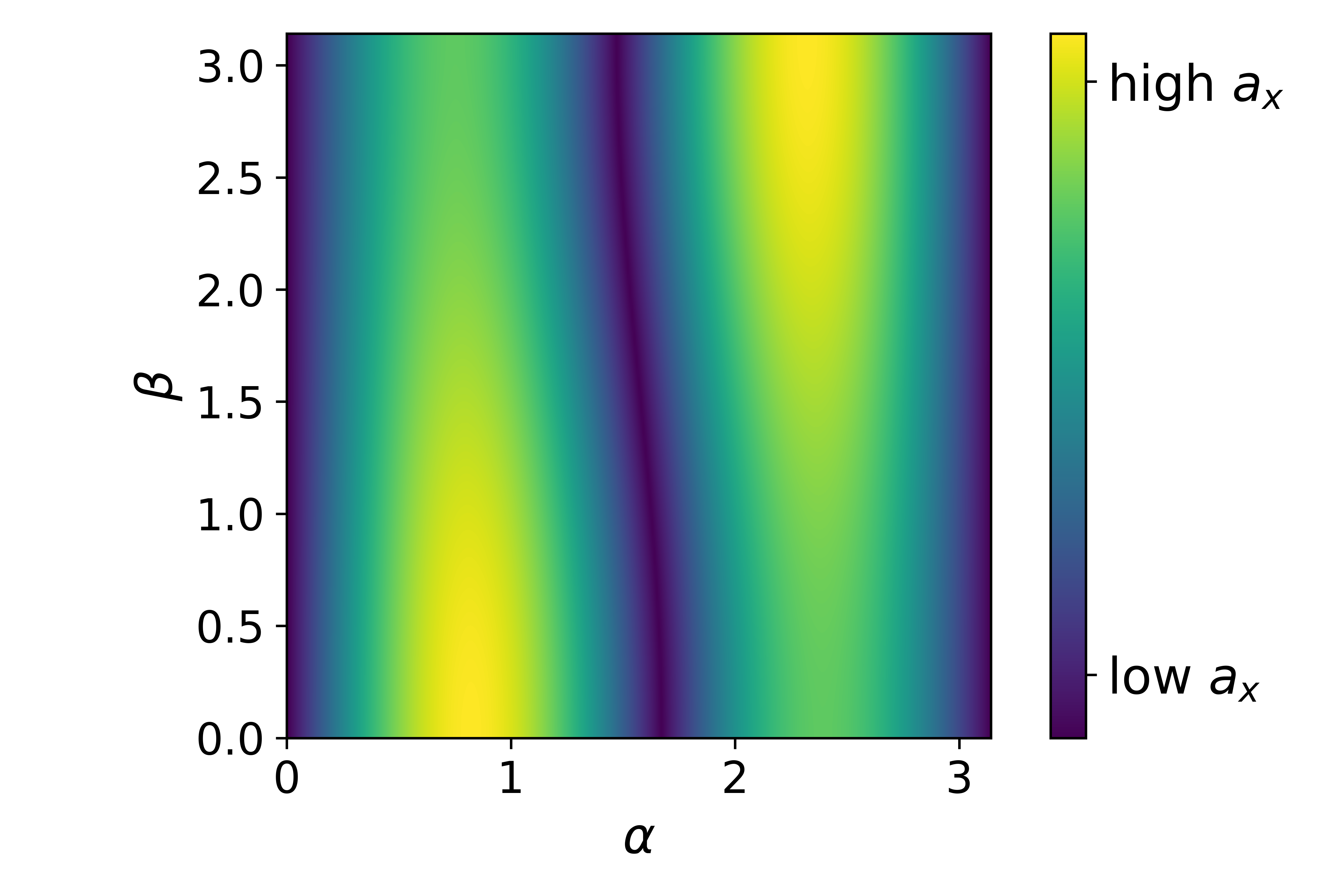}
  \caption{For $vt/b = 0.1$ the absolute value of the acceleration for the dipole case (Eq.~\eqref{eq:acc-angles-neutral} without the $\cos(\gamma)$-term) as a function of the projection angles $\alpha, \beta \in [0,\pi]$.}
  \label{fig:dipole-dipole-angles}
\end{minipage}%
\hfill
\end{figure*}

For a neutral interferometer there are the additional angles $\gamma$ and $\theta_0$ that give the projection in the $\hat{z}$-axis, see Eq.~\eqref{eq:second-proj-angles}~\footnote{
Note that for the induced dipole-charge interaction, these angles are not relevant, and the result for the projection angles for this interaction is the same as discussed in the previous section for the charged  interferometer case.
}.
They are relevant because the interferometer dipole has been chosen to align with the $\hat{z}$-axis.

Generically, we write the acceleration in the $\hat{x}$-direction from the dipole-charge and dipole-dipole interactions as:
\begin{equation}\label{eq:acc-angles-neutral}
    a(t) = \abs{C \frac{\cos(\alpha) + (vt/b) \cos(\beta)}{(1+(vt/b)^2)^{5/2}} \left( \cos(\theta_0) + (vt/b) \cos(\gamma) \right)}
\end{equation}
with $C$ a constant defined by the interaction.
The angles $\alpha$ and $\theta_0$ are related in the sense that if $\alpha=0$, it means that $b = x_0$, and therefore $\theta_0 = \pi/2$ such that $b_z = 0$.
Similarly the angles $\beta$ and $\gamma$ are related such that if $\beta = 0$, $v = v_x$ and thus $\gamma=\pi/2$, such that $v_z = 0$.
Therefore, we define the dependency as 
\begin{equation}
    \cos(\theta_0) = \sqrt{1 - \cos^2(\alpha)} , \, \, \, 
    \cos(\gamma) = \sqrt{1 - \cos^2(\beta)} .
\end{equation}
This follows from Pythagoras's theorem $ v2 = v_x2 + v_y2 + v_z2$, where we have set $v_y =0$ to maximise the acceleration.

The resulting plots that show the acceleration of Eq.~\eqref{eq:acc-angles-neutral} as a function of $\alpha$, $\beta$ for arbitrary $C$ are shown in Figs.~\ref{fig:neutral-angles-01}-\ref{fig:neutral-angles-10}.
The figures are somewhat similar to Figs.~\ref{fig:charged-angles-01}-\ref{fig:charged-angles-10}, but one can see the extra dependence on the projection angles. 
From the figures, we can conclude that to maximise the acceleration generically, one can take the projection angles $\alpha=\beta=\pi/4, 3\pi/4$, which means that $\theta_0 = \gamma = \pi/4, 3\pi/4$.

In the case of a dipole-dipole interaction, the assumption that the particle has low velocity such that $\boldsymbol{r}(t)\cdot\boldsymbol{d}$ is approximately constant in time removes the dependence on $\gamma$.
The resulting plot looks very similar to Fig.~\ref{fig:neutral-angles-01} and shows that the acceleration is maximal for $\alpha = \pi/4$, $\beta = 0$ and for $\alpha = 3\pi/4$ and $\beta = \pi$, see Fig.~\ref{fig:dipole-dipole-angles}.

\section{Approximations of the averaged dephasing}\label{app:approximations}
In section~\ref{sec:applications}, the averaged acceleration noise PSD of Eq.~\eqref{eq:Saa2} was used with the probability distributions over the impact parameter and projection angles (chosen to be a uniform distribution), and the velocity (determined to be a Maxwell-Boltzmann distribution).
For small temperatures, the width of the Maxwell-Boltzmann distribution becomes small and has most probable velocity $\bar{v} = \sqrt{2 k_b T_\text{gas}/m_\text{gas}}$.
We, therefore, consider the Maxwell-Boltzmann distribution to be a Dirac-delta distribution at $v=\bar{v}$, which holds approximately for $T_\text{gas}\leq 0.1\,\si{\milli\kelvin}$.
Since there is an integration over $v$ and $b$, taking $T=b/v$ as the characteristic timescale does not make sense anymore, and instead, we take $T=10\tau$ such that the timescale of the noise is larger than the interferometer time and thus Eq.~\eqref{eq:Teq} holds.
To simplify the calculation, we make further assumptions such that the Bessel functions can be approximated. 
The modified Bessel functions of the second kind arise from the Fourier transform of the acceleration, see Appendix~\ref{app:detail_calc} and more concretely Eqs.~\eqref{eq:Saadc-app},~\eqref{eq:Saadd-app} for the relevant acceleration PSDs used in the QGEM case and Eq.~\eqref{eq:Saacc-app} for the PSD used in the CNOT case.
These PSDs are used in this section in Eq.~\eqref{eq:Saa2} to find the averaged acceleration noise PSD given the distributions in Eqs.~\eqref{eq:1}-\eqref{eq:4}.
The Bessel functions that are contained in these PSDs can be approximated with the expansion in Eqs.~\eqref{eq:bessel1},~\eqref{eq:bessel2}, dependent on the parameters of the experiment.

\subsection{Approximations involved in the QGEM setup}

We approximate the integration over $\omega$ to get the total dephasing by substituting the dominant mode $\omega=\omega_\text{min}$.
This can be understood physically as considering the dephasing at the longest experimental time.
Mathematically, the transfer function (see Eq.~\eqref{eq:transferfunction}) shows strong decay for increasing $\omega$, showing that the minimal value is dominant and that the approximation is applicable due to the fast decline.
For $\omega=\omega_\text{min}=2\pi/\tau=2\pi\,\si{\hertz}$, the argument of the Bessel function is (see Eqs.~\eqref{eq:gamma-dc(ind)},~\eqref{eq:gamma-dc(per)} and~\eqref{eq:gamma-dd}):
\begin{equation}
    \frac{2\pi R}{\sqrt{2 k_b T_\text{gas}/m_\text{gas}}}\leq \frac{b \omega_\text{min}}{\bar{v}} \leq \frac{2\pi L}{\sqrt{2 k_b T_\text{gas}/m_\text{gas}}} \, ,
\end{equation}
where $m_\text{gas}$ denotes the mass of air molecules and we set $T_\text{gas} = 10^{-4}\,\si{\kelvin}$. For these parameters we can also approximate the Maxwell-Boltzmann distribution as discussed in the main text.
For a spherical diamond of mass $10^{-15}\,\si{\kilogram}$ (i.e. a radius of $\sim 1\,\si{\micro\metre}$) and an experimental chamber of $L=0.01\,\si{\metre}$, which are parameters applicable to the QGEM experiment, the argument of the Bessel function is small in the limit which we are considering in our computation: 
\begin{equation}
    b\omega_\text{min}/\bar{v} < 1 \text{ for } b\in[R,L].
\end{equation}
Therefore, we approximate the modified Bessel functions of the second kind as follows:
\begin{equation}\label{eq:bessel1}
    K_n(u) \approx \frac{\Gamma(n)}{2} \left(\frac{2}{u}\right)^n  \text{  for } 0 <\abs{u} \leq \sqrt{n+1} \text{ and } n>0 \, .
\end{equation}
With the approximate Bessel function, we find the dephasing of the dominant mode presented in Fig.~\ref{fig:QGEM} as a function of the number density $n_v=N/V$ with $V=L^3$.

\subsection{Approximations involved in the CNOT setup}
For $\omega=\omega_\text{min}=2\pi/\tau=2\pi*10^{-6} \,\si{\hertz}$, the argument of the Bessel function is given by:
\begin{equation}
    \frac{2\pi b_\text{min}}{\sqrt{2 k_b T_\text{gas}/m_\text{gas}}}\leq \frac{b \omega}{\bar{v}} 
    \, ,
\end{equation}
where $m_\text{gas}$ is the mass of air molecules, and $T_\text{gas} = 10^{-4}\,\si{\kelvin}$ such that the Maxwell-Boltzmann distribution can be approximated as discussed in the main text.
We find that the  argument of the Bessel function is large:
\begin{equation}
b \omega/\bar{v} > 1 \text{ for } b\in[10^3 R,L] \text{ and } \omega\in[\omega_\text{min},\infty],
\end{equation}
where we recall $L=0.01\,\si{\metre}$.
In order to proceeded analytically, we assume the minimal distance $b_\text{min} = 10^3 R \sim 10^{-7}\,\si{\metre}$, such that the modified Bessel functions of the second kind can be approximated via:
\begin{equation}\label{eq:bessel2}
    K_n(u) \approx e^{-u} \left(\left(\frac{\pi}{2 u}\right)^{1/2} + \ldots \right) \qq{} \text{for } u \to \infty \, .
\end{equation}
With the approximate Bessel function, we find the dephasing after integrating over $\omega$. 
This was presented in Fig.~\ref{fig:CNOT} as a function of the number density $n_v=N/V$ with $V=L^3$.

\end{document}